\documentclass[structabstract]{aa} %paper
\usepackage{CJK}
\usepackage{booktabs}
\usepackage{amsmath}
\usepackage{graphicx}
\usepackage{txfonts}
\usepackage{natbib}
\usepackage{longtable}
\usepackage{lscape}
\usepackage{multirow}
\usepackage{pdfpages}
\usepackage[colorlinks=true, allcolors=blue]{hyperref}

\bibpunct{(}{)}{;}{a}{}{,}

\newcommand{\kms}{km\,s$^{-1}$}

\newcommand{\Msun}{$M_{\odot}$}

\newcommand{\nodata}{...}
\newcommand{\accunit}{$M_{\odot}$~yr$^{-1}$}

\begin{document}

%\title{An ALCHEMI study of HCNH$^{+}$ as a tracer of low-density regions: An abundance enhancement in the heart of the starburst galaxy NGC~253}

\title{Shock-induced HCNH$^{+}$ abundance enhancement in the heart of the starburst galaxy NGC~253 unveiled by ALCHEMI}

\author{Y.~Gong\inst{1,2}, C.~Henkel\inst{2,3}, C.~T.~Bop\inst{4,5}, J. G. Mangum\inst{6}, E. Behrens\inst{6,7}, F.~J.~Du\inst{1}, S.~B.~Zhang\inst{1}, S. Martin\inst{8,9}, K.~M. Menten\inst{2}, N.~Harada\inst{10,11,12}, M.~Bouvier\inst{13}, X.~D.~Tang\inst{3,14}, K.~Tanaka\inst{15}, S.~Viti\inst{13}, Y.~T.~Yan\inst{2}, W.~Yang\inst{16}, R.~Q.~Mao\inst{1}, D.~H.~Quan\inst{17}}
\offprints{Y.~Gong, \email{ygong@pmo.ac.cn}}

\institute{
Purple Mountain Observatory, and Key Laboratory of Radio Astronomy, Chinese Academy of Sciences, 10 Yuanhua Road,
Nanjing 210023, PR China
\and
Max-Planck-Institut f{\"u}r Radioastronomie, Auf dem H{\"u}gel 69, D-53121 Bonn, Germany
\and
Xinjiang Astronomical Observatory, Chinese Academy of Sciences, 830011 Urumqi, PR China 
\and
Univ Rennes, CNRS, IPR (Institut de Physique de Rennes) - UMR 6251, F-35000 Rennes, France
\and 
Nantes Universit{\'e}, CNRS, CEISAM, UMR 6230, F-44000 Nantes, France
\and
National Radio Astronomy Observatory, 520 Edgemont Road, Charlottesville, VA  22903-2475, USA
\and
Department of Astronomy, University of Virginia, P. O. Box 400325, 530 McCormick Road, Charlottesville, VA 22904-4325, USA
\and 
European Southern Observatory, Alonso de C{\'o}rdova, 3107, Vitacura, Santiago 763-0355, Chile
\and
Joint ALMA Observatory, Alonso de C{\'o}rdova, 3107, Vitacura, Santiago 763-0355, Chile
\and
National Astronomical Observatory of Japan, 2-21-1 Osawa, Mitaka, Tokyo 181-8588, Japan
\and 
Institute of Astronomy and Astrophysics, Academia Sinica, 11F of AS/NTU Astronomy-Mathematics Building, No.1, Sec. 4, Roosevelt Road, Taipei 10617, Taiwan
\and 
Department of Astronomy, School of Science, The Graduate University for Advanced Studies (SOKENDAI), 2-21-1 Osawa, Mitaka, Tokyo 181-1855, Japan
\and
Leiden Observatory, Leiden University, PO Box 9513, 2300 RA Leiden, The Netherlands
\and
Key Laboratory of Radio Astronomy, Chinese Academy of Sciences, 830011 Urumqi, PR China
\and 
Department of Physics, Faculty of Science and Technology, Keio University, 3-14-1 Hiyoshi, Yokohama, Kanagawa 223-8522, Japan
\and
School of Astronomy \& Space Science, Nanjing University, 163 Xianlin Avenue, Nanjing 210023, People's Republic of China 
\and 
Research Center for Intelligent Computing Platforms, Zhejiang Laboratory, Hangzhou 311100, PR China
}

\date{Received date ; accepted date}

\abstract
{Understanding the chemistry of molecular clouds is pivotal to elucidate star formation and galaxy evolution. As one of the important molecular ions, HCNH$^{+}$ plays an important role in this chemistry. Yet, its behavior and significance under extreme conditions, such as in the central molecular zones (CMZs) of external galaxies, are still largely unexplored.}
{We aim to reveal the physical and chemical properties of the CMZ in the starburst galaxy NGC~253 with multiple HCNH$^{+}$ transitions to shed light on the molecule's behavior under the extreme physical conditions of a starburst.}
{We employ molecular line data including results for four rotational transitions of HCNH$^{+}$ from the ALMA Comprehensive High-resolution Extragalactic Molecular Inventory (ALCHEMI) large program to investigate underlying physical and chemical processes.}
{Despite weak intensities, HCNH$^{+}$ emission is widespread throughout NGC~253's CMZ, which suggests that this molecular ion can effectively trace large-scale structures within molecular clouds. Using the quantum mechanical coupled states approximation, we computed rate coefficients for collisions of HCNH$^{+}$ with {\it para}-H$_2$ and {\it ortho}-H$_2$ at kinetic temperatures up to 500 K. Using these coefficients in a non-LTE modeling framework and employing a Monte Carlo Markov chain analysis, we find that HCNH$^{+}$ emission originates from regions with H$_2$ number densities of $\sim 10^{2.80}-10^{3.55}$~cm$^{-3}$, establishing HCNH$^{+}$ as a tracer of low-density environments. Our analysis reveals that most of the HCNH$^{+}$ abundances in the CMZ of NGC~253 are higher than all reported values in the Milky Way. We performed static, photodissociation region (PDR), and shock modeling, and found that recurrent shocks could potentially account for the elevated HCNH$^{+}$ abundances observed in this CMZ.}
{We propose that the unexpectedly high HCNH$^{+}$ abundances may result from chemical enhancement, primarily driven by the elevated gas temperatures and cosmic ray ionization rates of shocked, low-density gas in the nuclear starburst regions of NGC~253.}
\titlerunning{HCNH$^{+}$ in NGC~253}
\authorrunning{Y. Gong et al.}
\maketitle
%________________________________________________________________
\section{Introduction}
Understanding the chemistry of molecular clouds in various environments affected by different physical and chemical boundary conditions is pivotal to gain insights into star formation and galaxy evolution. Molecular ions are crucial in shaping the chemistry of molecular clouds through ion-neutral reactions \citep[e.g.,][]{2013ChRv..113.8710A}. As the simplest protonated nitrile, HCNH$^{+}$ was first detected in the interstellar medium (ISM) almost four decades ago \citep{1986ApJ...302L..31Z}. Previous studies suggest that the abundance of HCNH$^{+}$ varies with different stages of star formation \citep{2021A&A...651A..94F,2023A&A...679A..39G}, indicating its potential as an evolutionary signpost. Numerical simulations further propose that HCNH$^{+}$, along with its chemically related species HCN, could help probe ambipolar diffusion in star formation regions \citep{2023MNRAS.521.5087T}. This molecular ion has also been suggested to play an important role in the formation of formamide \citep[NH$_2$CHO;][]{Kerkeni2023} which is a key prebiotic precursor \citep{2019ESC.....3.2122L}. Despite its importance in star formation and astrochemistry, HCNH$^{+}$ has received comparatively less attention than more commonly studied molecular ions like HCO$^{+}$ and N$_{2}$H$^{+}$. 
Because of its small dipole moment of 0.29~D \citep{1986CPL...124..382B} and the resultant low H$_{2}$ critical densities \citep{2023A&A...679A..39G}, HCNH$^{+}$ is expected to trace large-scale structures within molecular clouds across galaxies. However, previous observations of HCNH$^{+}$ have been predominantly confined to Galactic molecular clouds on linear scales of $\lesssim$10~pc \citep[e.g.,][]{1986ApJ...302L..31Z,1991A&A...247..487S,2017MNRAS.470.3194Q,2023A&A...679A..39G}, leaving its behavior on larger scales and its role under extreme conditions largely unexplored.

At a distance of $\sim$3.5~Mpc \citep{2003A&A...404...93K,2005MNRAS.361..330R}, the Sculptor galaxy NGC~253 stands as one of the nearest and most prominent starburst galaxies. This archetypal starburst galaxy boasts a total star formation rate of $\sim$4.2~\accunit\,\citep{2003AJ....126.1607S} with $\sim$2~\accunit\,concentrated in the nuclear starburst region, its CMZ \citep{2015MNRAS.450L..80B,2015ApJ...801...25L}. The energetic starburst activity drives large-scale galactic outﬂows \citep[e.g.,][]{1985ApJ...299..312T,2000AJ....120.2965S,2013Natur.499..450B,2024A&A...687A..43B}. 
The starburst is fueled by inflowing gas from a bar \citep{2014A&A...567A..86I} and is characterized by the central molecular zone (CMZ) which harbors a total molecular gas mass of $\sim 10^{8}$~\Msun\,\citep[e.g.,][]{2011ApJ...735...19S}. The CMZ is significantly influenced by starburst feedback \citep[e.g.,][]{2021ApJ...923...24H}, and has proven to be a rich reservoir of molecular species \citep[e.g.,][]{2006ApJS..164..450M,2015A&A...579A.101A,2021A&A...656A..46M,2024ApJS..271...38H}. Recent studies have highlighted elevated cosmic-ray ionization rates \citep[e.g.,][]{2021ApJ...923...24H,2022ApJ...931...89H,2022ApJ...939..119B,2024ApJ...977...38B,2024MNRAS.531.2930P}, large-scale shocks \citep[e.g.,][]{2022A&A...663A..33H,2022ApJ...938...80H,2023A&A...675A.151H}, and a large amount of warm or hot  molecular gas \citep[e.g.,][]{2019ApJ...871..170M,2020MNRAS.491.4573R,2023A&A...675A.151H,2024ApJ...961...18T,2024A&A...689A..64B} within the CMZ. These extreme conditions make the CMZ of NGC~253 an exceptional target for investigating the behavior of HCNH$^{+}$ under extreme physical environments. 

Despite its potential, the weak intensities of its rotational transitions pose significant challenges for mapping its large-scale distribution of  HCNH$^{+}$ in galaxies. Thanks to the unprecedented sensitivity of the ALMA Comprehensive High-resolution Extragalactic Molecular Inventory (ALCHEMI) survey \citep{2021A&A...656A..46M}, HCNH$^{+}$ was successfully discovered in the CMZ of NGC~253 \citep{2024ApJS..271...38H} where only one single HCNH$^{+}$ transition was reported because of the selection criteria of the study. Recent advancements in the calculation of excitation rate coefficients for HCNH$^{+}$ collisions with H$_{2}$ enable the modeling of multiple HCNH$^{+}$ transitions to determine physical conditions within molecular clouds \citep{2023JChPh.158g4304B,2024A&A...681L..19B}. However, such modeling has yet to be applied in an extragalactic environment. In this study, we aim to investigate physical and chemical properties of the CMZ of NGC~253 with multiple transitions of HCNH$^{+}$, shedding light on its behavior under the extreme starburst physical conditions.

We describe our observations in Sect.~\ref{Sec:obs}. Our findings are presented in Sect.~\ref{Sec:res}, with discussions of the results in Sect.~\ref{Sec:dis}. The summary and conclusions are outlined in Sect.~\ref{Sec:sum}.

\section{Observations and data reduction}\label{Sec:obs}
The data used in this study are directly taken from the ALCHEMI survey \citep[co-PIs.: S.~Mart{\'i}n, N.~Harada, and J.~Mangum;][]{2021A&A...656A..46M}.
A comprehensive description of the survey has already been provided in \citet{2021A&A...656A..46M}, so we only offer a brief summary here. The ALCHEMI survey (project codes: 2017.1.00161.L, 2018.1.00162.S) spans ALMA Bands 3--7, covering frequencies from 84 to 375 GHz, excluding regions affected by atmospheric absorption features. The spectroscopic information was taken from the entry in the Cologne Database for Molecular Spectroscopy \citep[CDMS\footnote{https://cdms.astro.uni-koeln.de/},][]{2016JMoSp.327...95E}, which is based on the laboratory results of \citet{2024JChPh.160g1101S}. HCNH$^{+}$ (2--1), HCNH$^{+}$ (3--2), HCNH$^{+}$ (4--3), and HCNH$^{+}$ (5--4) have rest frequencies at 148221.450(17)~MHz, 222329.277(8)~MHz, 296433.652(17)~MHz, and 370533.375(23)~MHz, respectively, where uncertainties in the last digits are given in parentheses. These four HCNH$^{+}$ rotational transitions fall within the frequency range covered by the ALCHEMI survey. The phase center of these observations is ($\alpha_{\rm J2000}$, $\delta_{\rm J2000}$)=(00$^{\rm h}$47$^{\rm m}$33\rlap{.}$^{\rm s}$28, $-$25$^{\circ}$17$^{\prime}$17\rlap{.}$^{\prime\prime}$6). 
The surveyed area of 50\arcsec$\times$20\arcsec\, encompasses the entire CMZ of NGC~253. All data products have been convolved to a common circular beam of 1\rlap{.}\arcsec6 (i.e., 27~pc). 
%KMM: For the high frequency lines, the images should be strongly affected by primary beam attenuation. Is this accounted for?
%This should be described!
In this work, we mainly use the transitions of HCNH$^{+}$ for our analysis, and their maximum recoverable scales of $\gtrsim$11\arcsec\,was achieved by combining the 12 m Array and Atacama Compact Array data. The data cubes were extracted from the ALCHEMI continuum-subtracted FITS cubes using the SpectralCube\footnote{\url{https://spectral-cube.readthedocs.io/en/latest/}} python package. The observed flux densities were converted to the brightness temperature scale using the Rayleigh-Jeans formula. The absolute flux calibration uncertainties were found to be $\lesssim 10$\% \citep{2021A&A...656A..46M}. The typical rms noise levels are about 4--20 mK at a channel width of 10~\kms. The velocity scale has been calibrated with respect to the local standard of rest (LSR) in this work.
%Their rest frequencies are taken from the Cologne Database for Molecular Spectroscopy \citep[CDMS\footnote{https://cdms.astro.uni-koeln.de/},][]{2016JMoSp.327...95E}.

\section{Results}\label{Sec:res}
\subsection{Spatial distribution and spectra}
\begin{figure}[!htbp]
\centering
\includegraphics[width = 0.49 \textwidth]{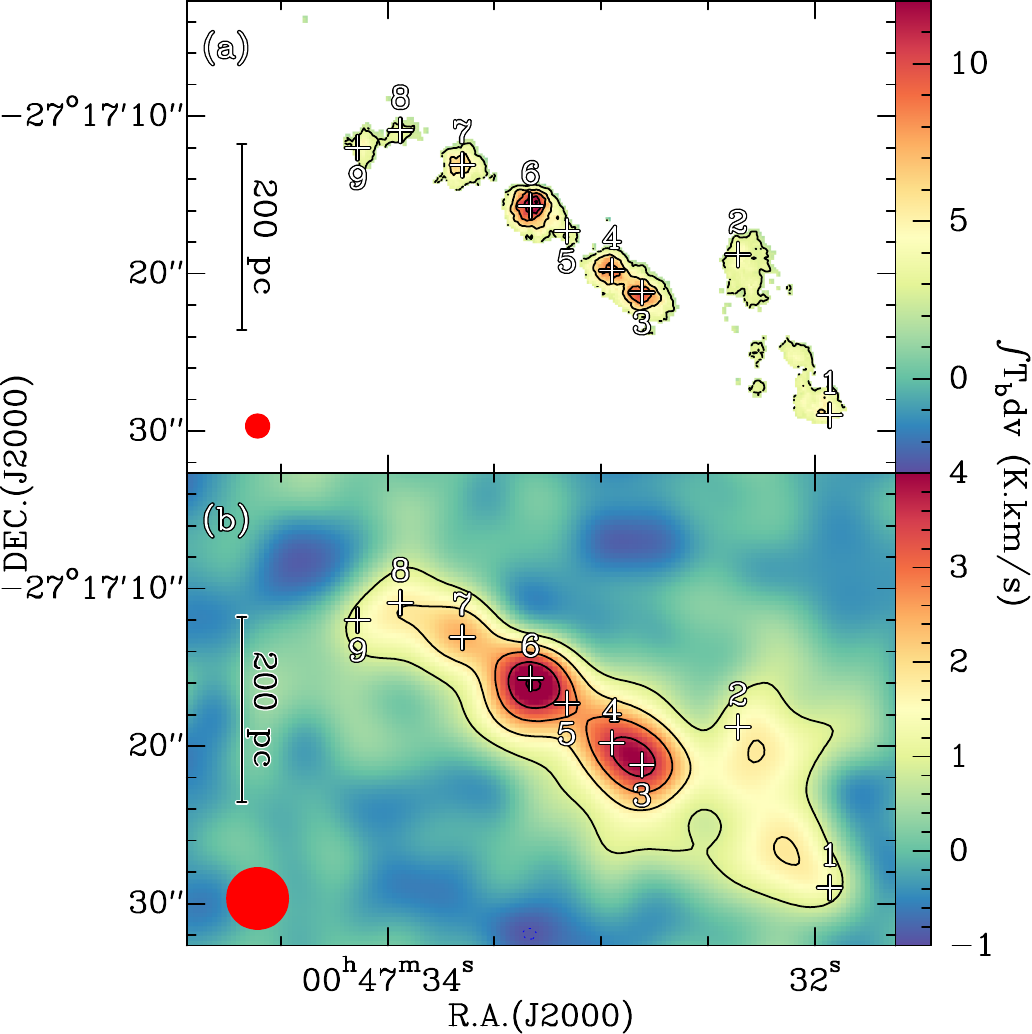}
\caption{{Distribution of HCNH$^{+}$ (2--1) in the CMZ of NGC~253. (a) HCNH$^{+}$ (2--1) integrated intensity map (taken from \citealt{2024ApJS..271...38H}),  clipped at the 3$\sigma$ level. The contours start from 2.7~K~\kms\,(3$\sigma$) and increase by 2.7~K~\kms. (b) HCNH$^{+}$ (2--1) integrated intensity map that has been convolved to a circular beam of 4\arcsec. The smoothed image manifests the large-scale structures traced by HCNH$^{+}$ transitions. The contours start from 0.9~K~\kms\,(3$\sigma$) and increase by 0.9~K~\kms. In each panel, the beam is shown in the lower-left corner, and the pluses indicate the positions of the nine most prominent GMCs.}\label{Fig:hcnh+_m0}}
\end{figure}

Figure~\ref{Fig:hcnh+_m0} illustrates the elongated distribution of HCNH$^{+}$ (2--1) emission within the CMZ of NGC~253, with the nine most prominent giant molecular clouds (GMCs) labeled \citep[see Table~1 in][]{2024ApJS..271...38H}. Figure~\ref{Fig:hcnh+_m0}a is adapted from Fig.~21 in \citet{2024ApJS..271...38H}, while Fig.~\ref{Fig:hcnh+_m0}b shows the integrated intensity map convolved to a circular beam of 4\arcsec\ FWHM,without sigma clipping. Although Fig.~\ref{Fig:hcnh+_m0}b does not recover the large-scale emission filtered out by the ALMA array, the smoothed image demonstrates that HCNH$^{+}$ emission is more widespread than previously thought, with a radial extent of $\sim$300~pc in the CMZ. This is because the sensitivity is not high enough in Fig.~\ref{Fig:hcnh+_m0}a to detect the weak signals seen in Fig.~\ref{Fig:hcnh+_m0}b. Especially, the emission peaks are largely offset with respect to the positions of GMC1 and GMC2 in the low-angular-resolution image (see Fig.~\ref{Fig:hcnh+_m0}b), which is caused by the smoothing process where low-density gas close to GMC1 and GMC2 substantially contributes to the HCNH$^{+}$ distribution. This behavior is expected due to the rather low upper-state energy levels and critical densities \citep[see Table~1 in][]{2023A&A...679A..39G}. Consequently, HCNH$^{+}$ emission effectively traces large-scale structures within molecular clouds. 

\begin{figure*}[!htbp]
\centering
\includegraphics[width = 0.95 \textwidth]{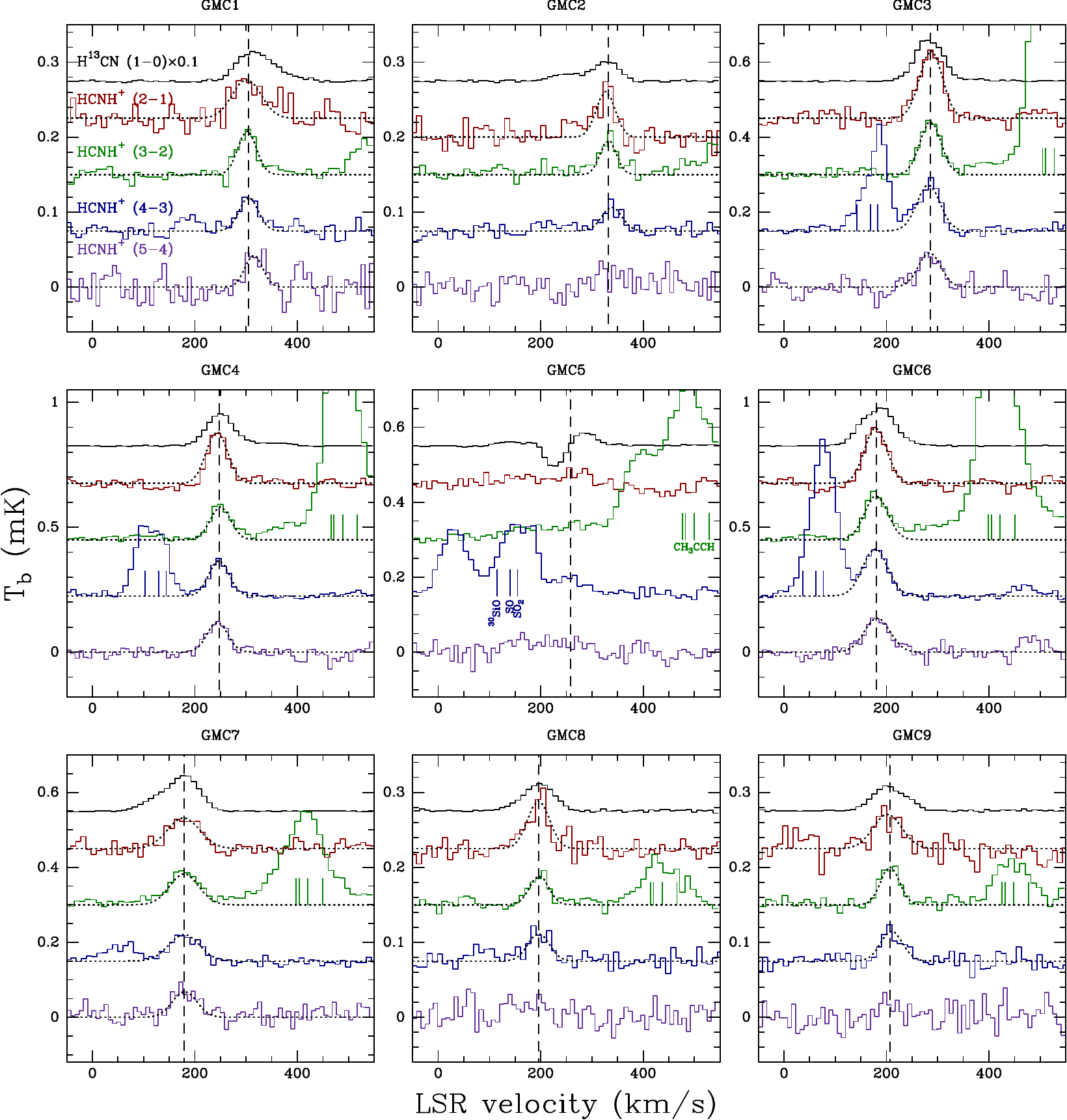}
\caption{{Observed H$^{13}$CN and HCNH$^{+}$ spectra (all with a common circular beam of 1\rlap{.}\arcsec6) of the nine GMCs indicated in Fig.~\ref{Fig:hcnh+_m0} and Table~\ref{Tab:obs}. H$^{13}$CN (1--0), HCNH$^{+}$ (2--1), HCNH$^{+}$ (3--2), HCNH$^{+}$ (4--3), and HCNH$^{+}$ (5--4) are indicated by black, red, green, blue, and purple lines, respectively. The fits to the HCNH$^{+}$ spectra are indicated by black dotted lines. H$^{13}$CN (1--0) spectra are scaled by a factor of 0.1 for better visualization. In each panel, the systemic LSR velocity is indicated by the vertical black dashed line. The potentially blended lines are marked by short vertical lines in the corresponding panels, with the associated species labeled in the central panel for GMC5.}\label{Fig:spec}}
\end{figure*}

Figure~\ref{Fig:spec} presents the observed HCNH$^{+}$ spectra for the nine GMCs (see their positions in Fig.~\ref{Fig:hcnh+_m0}). HCNH$^{+}$ is detected in eight of the GMCs, with at least two transitions in each GMC exceeding the 3$\sigma$ level, except for GMC5, the nucleus and kinematic center of the galaxy, which is associated with the brightest radio continuum emission \citep{1985ApJ...299L..77T}. The peak velocities of HCNH$^{+}$ transitions agree well with those of all other molecular species, supporting the assignment of these spectral features to HCNH$^{+}$. Additionally, there are spectral features near the HCNH$^{+}$ (3--2) and HCNH$^{+}$ (4--3) lines. The features close to HCNH$^{+}$ (3--2) correspond to multiple k-ladder transitions of CH$_{3}$CCH~($12_{k}-11_{k}$), while those near HCNH$^{+}$ (4--3) are likely due to the blend of SO~($6_{7}-5_{6}$) with a rest frequency of 296550.064(30)~MHz, the highly excited SO$_{2}$~($24_{4,20}-24_{3,21}$) with a rest frequency of 296535.4225(14)~MHz, and $^{30}$SiO (7--6) with a rest frequency of 296575.730(20)~MHz \citep{2016JMoSp.327...95E}. Thus, the classic integrated intensity maps over a specific velocity range (i.e., from 100 to 400~\kms) inevitably introduce contamination to the HCNH$^{+}$ (3--2) and (4--3) integrated intensity maps. However, the CH$_{3}$CCH features are redshifted by $\sim$240~\kms\,relative to HCNH$^{+}$ (3--2), and the blended lines around 296550~MHz are blueshifted by $\sim$100~\kms\,with respect to HCNH$^{+}$ (4--3), allowing us to isolate the observed properties of HCNH$^{+}$ transitions. To minimize this contamination, we applied a masking method before producing the integrated-intensity maps.

%Because HCNH$^{+}$ (3--2) and (4--3) are adjacent to other spectral lines (see Fig.~\ref{Fig:spec}), the commonly used zeroth-moment maps, integrated over a velocity range (i.e., 100--400~\kms), will introduce the contamination from neighboring transitions. 

Since the HCNH$^{+}$ (2--1) line is brighter than the other HCNH$^{+}$ transitions and less contaminated by emission from other species, we use emission in this line to create the mask both spatially and spectrally for the data cubes of all the HCNH$^{+}$ transitions. Our masking method\footnote{The masking script is available from \href{https://github.com/shbzhang/MaskingNGC253}{this link}.} is a modified version of that previously adopted by \citet{2024ApJS..271...38H}. We labeled the connected components above the 3$\sigma$ threshold in the three-dimensional ($\alpha_{2000}, \delta_{2000}$, $\varv_{\rm LSR}$) data cube of HCNH$^{+}$ (2--1) using a 26-connected neighborhood\footnote{In a single spectral plane, each pixel has 9 connected neighbors, including itself. In a data cube, each spectral plane has two adjacent spectral planes. Hence, each pixel has 27 connected neighbors in total. Excluding the pixel itself, 26-connected neighbors remain. If the central pixel has an intensity above 3$\sigma$ threshold, the 26-connected neighbors are included in the mask.} algorithm\footnote{\url{https://github.com/seung-lab/connected-components-3d}}. These labeled components were subsequently utilized as masks for all the HCNH$^{+}$ transitions. The mask areas were carefully expanded in the spatial dimensions to include surrounding emission and in the spectral dimension to capture line wings while minimizing contamination from other spectral lines. Furthermore, we manually masked the regions near HCNH$^{+}$ (4--3) emission toward GMC5 and GMC6 to mitigate potential contamination in the spectral axis. Data outside these masked regions were trimmed to focus the analysis on the relevant areas within the mask. 

\begin{figure*}[!htbp]
\centering
\includegraphics[width = 0.95 \textwidth]{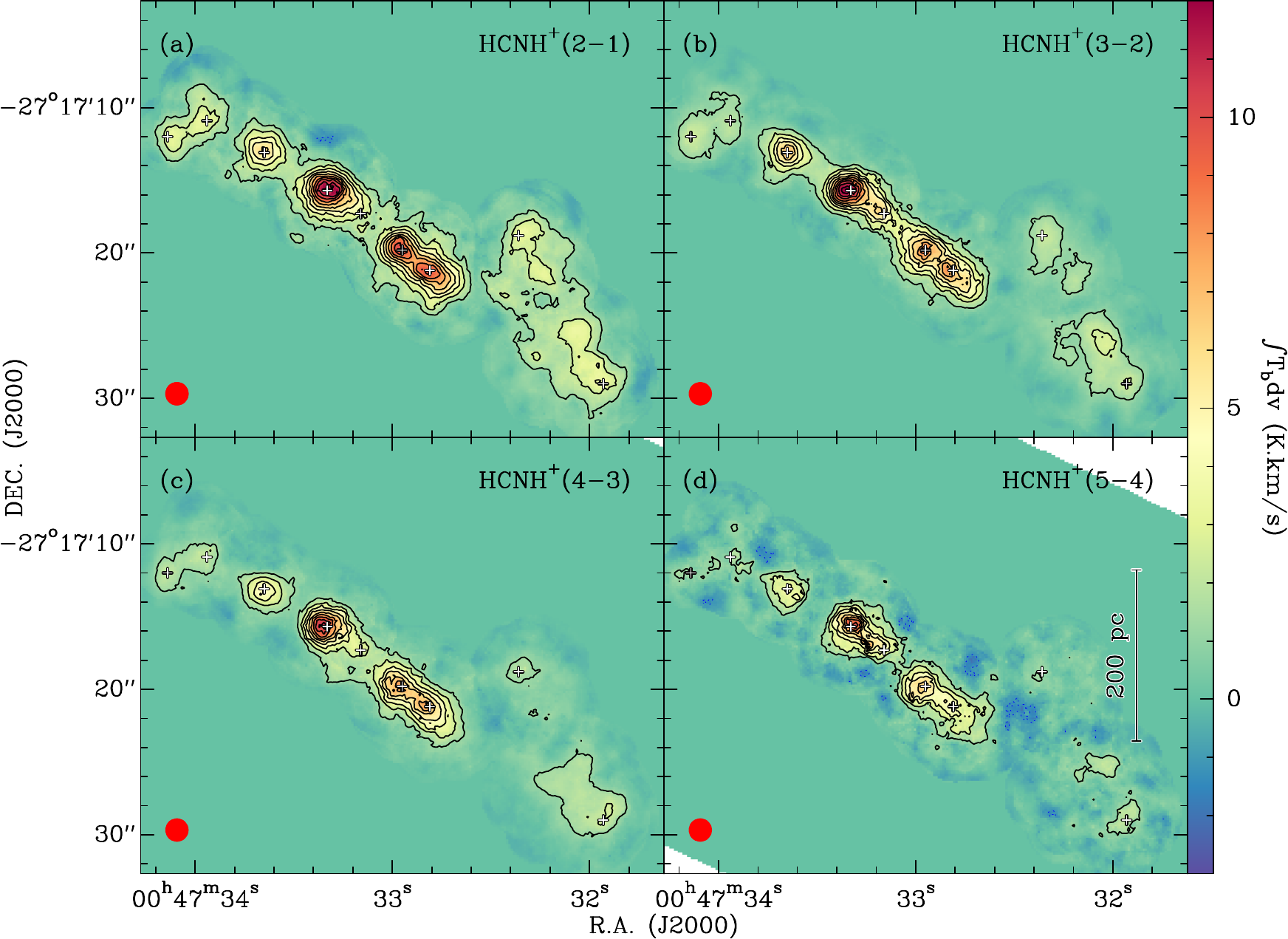}
\caption{{Distributions of HCNH$^{+}$ (2--1) (a), HCNH$^{+}$ (3--2) (b), HCNH$^{+}$ (4--3) (c), and HCNH$^{+}$ (5--4) (d) emission in the CMZ of NGC~253. All intensity maps are integrated from 100 to 400~\kms\, and share a common integrated-intensity scale indicated by the color bar. In each panel, the overlaid contours represent the integrated intensities of respective HCNH$^{+}$ transitions, starting at 1.2~K~\kms\,and increasing by 1.2~K~\kms. The positions of the nine GMCs are marked with plus symbols. In each panel, the beam is shown in the lower-left corner.}\label{Fig:4m0}}
\end{figure*}

Based on the trimmed data cubes\footnote{The data cubes will be made public online via \href{https://gongyan2444.github.io/}{this link}.}, we derived the integrated intensity distributions of the four HCNH$^{+}$ transitions over a range from 100~\kms\,to 400~\kms. The results are presented in Fig.~\ref{Fig:4m0}. The integrated intensities of Fig.~\ref{Fig:hcnh+_m0}a and Fig.~\ref{Fig:4m0}a are consistent within uncertainties, with Fig.~\ref{Fig:4m0}a showing a higher dynamical range and a more widespread distribution than Fig.~\ref{Fig:hcnh+_m0}a. This consistency suggests that our masking approach accurately reproduces the intrinsic integrated intensities and enhances the detection of low-intensity emissions.
Figure~\ref{Fig:4m0} not only confirms the widespread distribution of HCNH$^{+}$ emission but also shows that the four HCNH$^{+}$ transitions exhibit nearly identical spatial distributions, with emission peaking at GMC6. The main difference is the different integrated intensities of these transitions, with HCNH$^{+}$ (2--1) being the brightest. %The integrated intensities of HCNH$^{+}$ (2--1) are generally higher than those of the higher $J$ HCNH$^{+}$ transitions. 
This indicates that HCNH$^{+}$ transitions predominantly trace the same gas structures, albeit with varying intensities. While weak HCNH$^{+}$ emission appears to be present in GMC5 in Fig.~\ref{Fig:4m0}, its spectra lack sufficient signal-to-noise ratios in Fig.~\ref{Fig:spec}. Therefore, HCNH$^{+}$ emissions in GMC5 are excluded from further analysis.

\begin{table*}[!hbt]
\caption{Observational parameters of the selected GMCs.}\label{Tab:obs}
\small
\centering
\renewcommand{\arraystretch}{1.2}
\begin{tabular}{ccccccc}
\hline \hline
Target    & ($\alpha_{\rm J2000}$, $\delta_{\rm J2000}$) & Transition & $\varv_{\rm LSR}$ & $\Delta \varv_{\rm LSR}$ & $T_{\rm p}$ & $\int T_{\rm b}$d$\varv$   \\ 
          & ($^{\rm h}$:$^{\rm m}$:$^{\rm s}$, $^{\circ}$:$^{\prime}$:$^{\prime\prime}$) &                & (\kms)         & (\kms)  & (mK)         & (K~\kms)          \\ 
(1)       &                (2)                           & (3)        & (4)        & (5)            & (6)     & (7)                        \\
\hline
GMC1 & (00:47:31.93, $-$25:17:29.0) & HCNH$^{+}$ (2--1) & 301$\pm$4 & 66$\pm$9  & 52$\pm$12 & 3.63$\pm$0.45 \\
     & & HCNH$^{+}$ (3--2) & 302$\pm$4 & 42$\pm$9  & 59$\pm$6  & 2.66$\pm$0.47 \\
     & & HCNH$^{+}$ (4--3) & 305$\pm$3 & 45$\pm$8  & 43$\pm$4  & 2.03$\pm$0.29 \\
     & & HCNH$^{+}$ (5--4) & 315$\pm$6 & 40$\pm$13 & 43$\pm$8  & 1.83$\pm$0.54 \\
\hline     
GMC2 & (00:47:32.36, $-$25:17:18.8) & HCNH$^{+}$ (2--1) & 327$\pm$3 & 43$\pm$8  & 61$\pm$15 & 2.75$\pm$0.40 \\
     & & HCNH$^{+}$ (3--2) & 332$\pm$3 & 36$\pm$10 & 47$\pm$6  & 1.81$\pm$0.37 \\
     & & HCNH$^{+}$ (4--3) & 339$\pm$3 & 45$\pm$9  & 32$\pm$7  & 1.54$\pm$0.24 \\
     & & HCNH$^{+}$ (5--4) & \nodata   & \nodata   & $<$36     & $<$1.62 \\
\hline
GMC3 & (00:47:32.81, $-$25:17:21.2) & HCNH$^{+}$ (2--1) & 287$\pm$1 & 52$\pm$3  & 181$\pm$15 & 10.06$\pm$0.51\\
     & & HCNH$^{+}$ (3--2) & 286$\pm$1 & 48$\pm$2  & 143$\pm$6  & 7.30$\pm$0.30 \\
     & & HCNH$^{+}$ (4--3) & 282$\pm$1 & 53$\pm$3  & 123$\pm$17 & 6.92$\pm$0.30 \\
     & & HCNH$^{+}$ (5--4) & 284$\pm$3 & 51$\pm$8  & 92$\pm$13  & 5.02$\pm$0.64 \\
 \hline
GMC4 & (00:47:32.95, $-$25:17:19.8) & HCNH$^{+}$ (2--1) & 243$\pm$3 & 49$\pm$6  & 208$\pm$10 & 10.76$\pm$1.21\\
     & & HCNH$^{+}$ (3--2) & 250$\pm$1 & 46$\pm$2  & 133$\pm$11 & 6.54$\pm$0.25 \\
     & & HCNH$^{+}$ (4--3) & 247$\pm$1 & 46$\pm$2  & 143$\pm$9  & 6.96$\pm$0.20 \\
     & & HCNH$^{+}$ (5--4) & 244$\pm$2 & 47$\pm$4  & 120$\pm$13 & 6.00$\pm$0.40 \\ 
 \hline
GMC5 & (00:47:33.16, $-$25:17:17.3) & HCNH$^{+}$ (2--1) & \nodata   & \nodata   & $<$41  & $<$1.85\\
     & & HCNH$^{+}$ (3--2) & \nodata   & \nodata   & $<$38  & $<$1.71 \\
     & & HCNH$^{+}$ (4--3) & \nodata   & \nodata   & $<$42  & $<$3.24 \\
     & & HCNH$^{+}$ (5--4) & \nodata   & \nodata   & $<$54  & $<$2.43 \\   
 \hline
GMC6 & (00:47:33.33, $-$25:17:15.7) & HCNH$^{+}$ (2--1) & 179$\pm$1 & 51$\pm$3  & 225$\pm$11 & 12.14$\pm$0.57 \\
     & & HCNH$^{+}$ (3--2) & 180$\pm$1 & 53$\pm$2  & 177$\pm$14 & 10.05$\pm$0.31 \\
     & & HCNH$^{+}$ (4--3) & 178$\pm$1 & 57$\pm$2  & 190$\pm$6  & 11.66$\pm$0.36 \\
     & & HCNH$^{+}$ (5--4) & 182$\pm$2 & 59$\pm$6  & 137$\pm$13 & 8.63$\pm$0.71 \\
 \hline
GMC7 & (00:47:33.65, $-$25:17:13.1) & HCNH$^{+}$ (2--1) & 178$\pm$4 & 74$\pm$8  & 82$\pm$9   & 6.53$\pm$0.68 \\
     & & HCNH$^{+}$ (3--2) & 179$\pm$2 & 66$\pm$5  & 83$\pm$6   & 5.83$\pm$0.38 \\
     & & HCNH$^{+}$ (4--3) & 179$\pm$2 & 70$\pm$5  & 71$\pm$9   & 5.27$\pm$0.31 \\
     & & HCNH$^{+}$ (5--4) & 180$\pm$5 & 56$\pm$11 & 69$\pm$15  & 4.12$\pm$0.70 \\
\hline
GMC8 & (00:47:33.94, $-$25:17:10.9) & HCNH$^{+}$ (2--1) & 196$\pm$3 & 50$\pm$10 & 64$\pm$13  & 3.40$\pm$0.44 \\
     & & HCNH$^{+}$ (3--2) & 197$\pm$2 & 43$\pm$5  & 39$\pm$5   & 1.81$\pm$0.18 \\
     & & HCNH$^{+}$ (4--3) & 196$\pm$3 & 49$\pm$7  & 37$\pm$10  & 1.93$\pm$0.27 \\
     & & HCNH$^{+}$ (5--4) & \nodata   & \nodata   & $<$43      & $<$2.13       \\
\hline
GMC9 & (00:47:34.14, $-$25:17:12.0) & HCNH$^{+}$ (2--1) & 202$\pm$6 & 66$\pm$15 & 46$\pm$11  & 3.21$\pm$0.61 \\
     & & HCNH$^{+}$ (3--2) & 207$\pm$2 & 43$\pm$4  & 51$\pm$5   & 2.32$\pm$0.18 \\
     & & HCNH$^{+}$ (4--3) & 210$\pm$3 & 50$\pm$7  & 40$\pm$7   & 2.08$\pm$0.24 \\
     & & HCNH$^{+}$ (5--4) & \nodata   & \nodata   & $<$45      & $<$2.27       \\
\hline
\end{tabular}
\tablefoot{(1) Selected GMC. (2) Equatorial coordinate. (3) Transition. (4) Velocity centroid. (5) Full width at half-maximum line width. (6) Peak main beam brightness temperature. (7) Integrated intensity.}
\normalsize
\end{table*}

We fit the observed HCNH$^{+}$ transitions from the selected GMCs using a single-component Gaussian model, with the results presented in Table~\ref{Tab:obs}. 
The fitted velocity centroids range from $\sim$180~\kms\,to $\sim$330~\kms\ and full width at half-maximum (FWHM) line widths are in the range of 40--70~\kms. These values are consistent with previously reported velocities determined for other molecular tracers within 3$\sigma$ uncertainties \citep[e.g.,][]{2022A&A...659A.158H,2022A&A...663A..33H,2024A&A...689A..64B}. Based on the fitted results, we find that all HCNH$^{+}$ transitions have peak intensities of $\lesssim$200~mK, which are much lower than those of H$^{13}$CN (1-0) (see Fig.~\ref{Fig:spec}). Given the widespread distribution of HCNH$^{+}$ emission in the CMZ (see Fig.~\ref{Fig:hcnh+_m0}b), beam dilution effects might be negligible. Assuming an excitation temperature of 15~K (as discussed below) and neglecting beam dilution effects, we find that the low peak intensities correspond to optical depths of $<$0.1. Even accounting for a beam dilution factor as low as 0.1, the peak optical depths remain below 0.3. Therefore, we conclude that the HCNH$^{+}$ transitions are optically thin in the CMZ of NGC~253.

\subsection{LTE analysis}
\begin{figure*}[!htbp]
\centering
\includegraphics[width = 0.49 \textwidth]{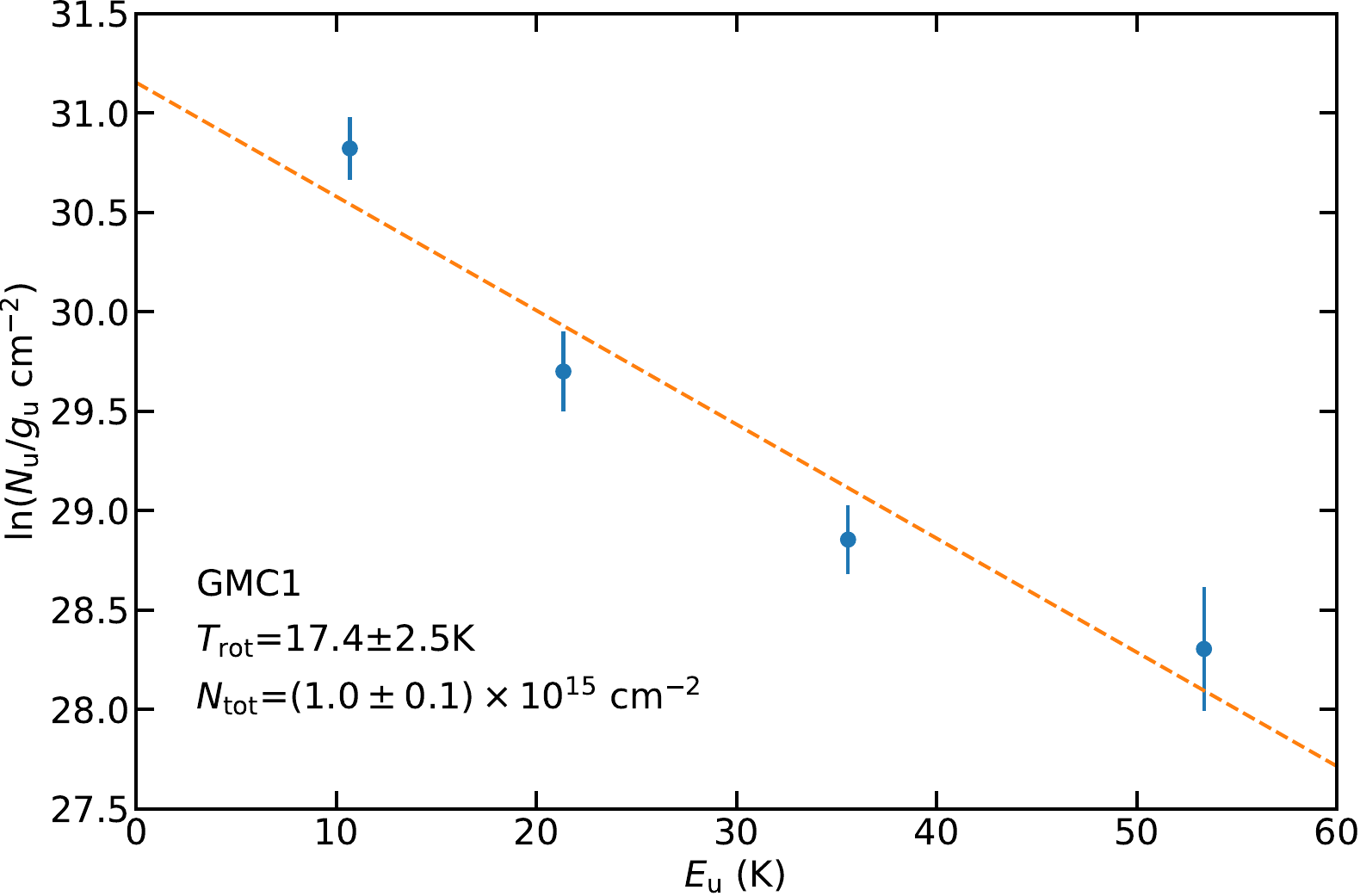}
\includegraphics[width = 0.49 \textwidth]{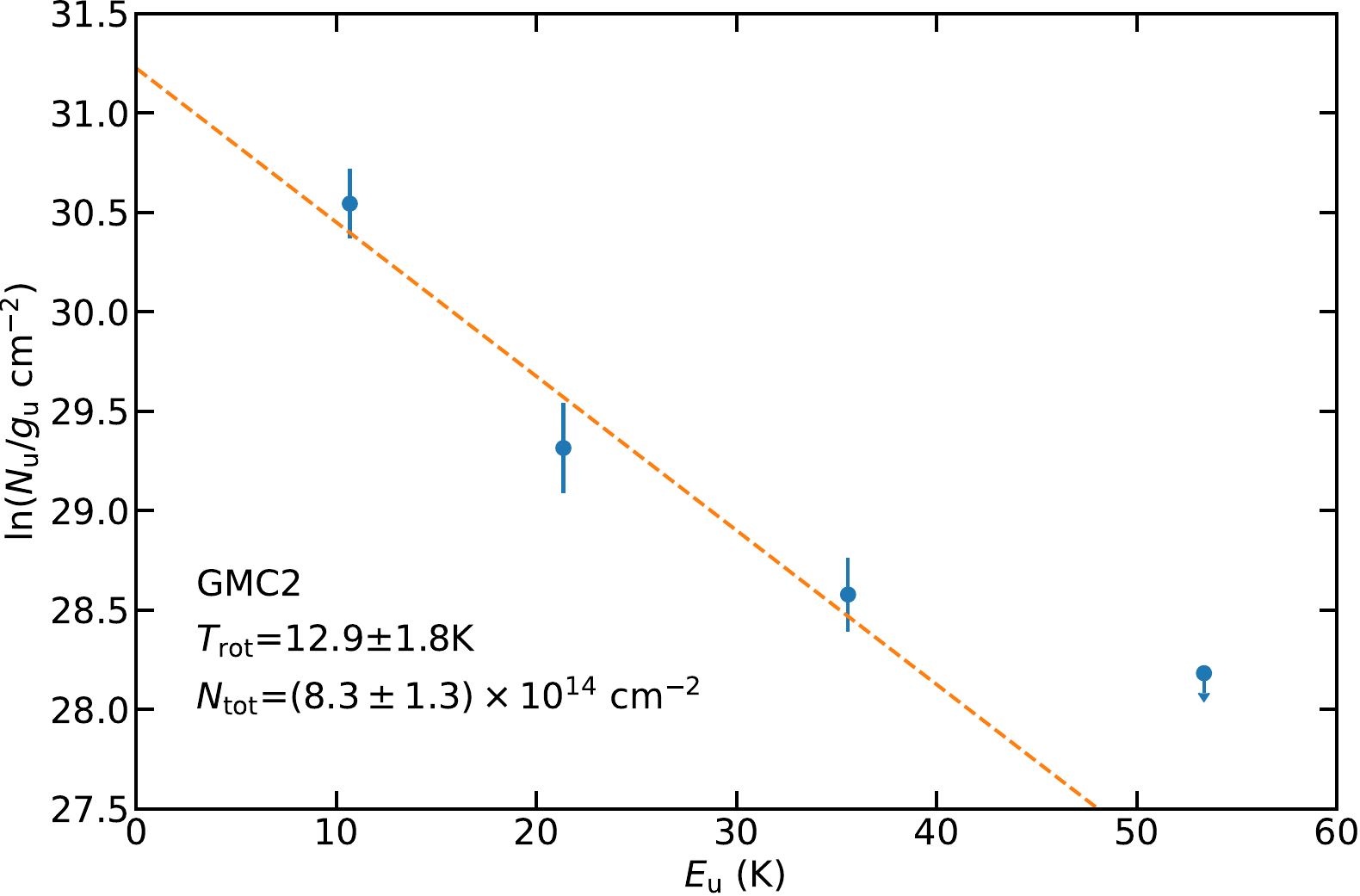}
\includegraphics[width = 0.49 \textwidth]{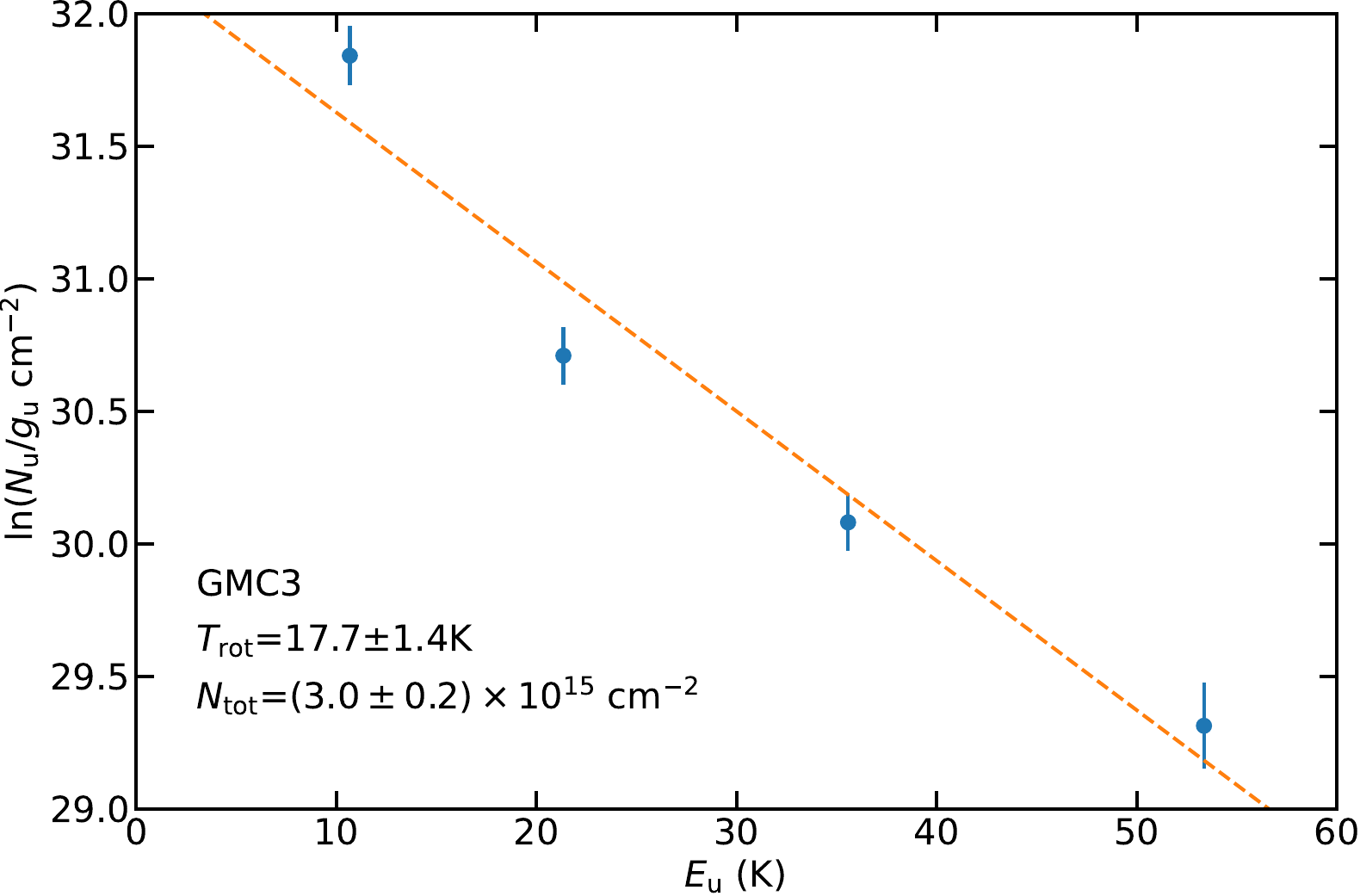}
\includegraphics[width = 0.49 \textwidth]{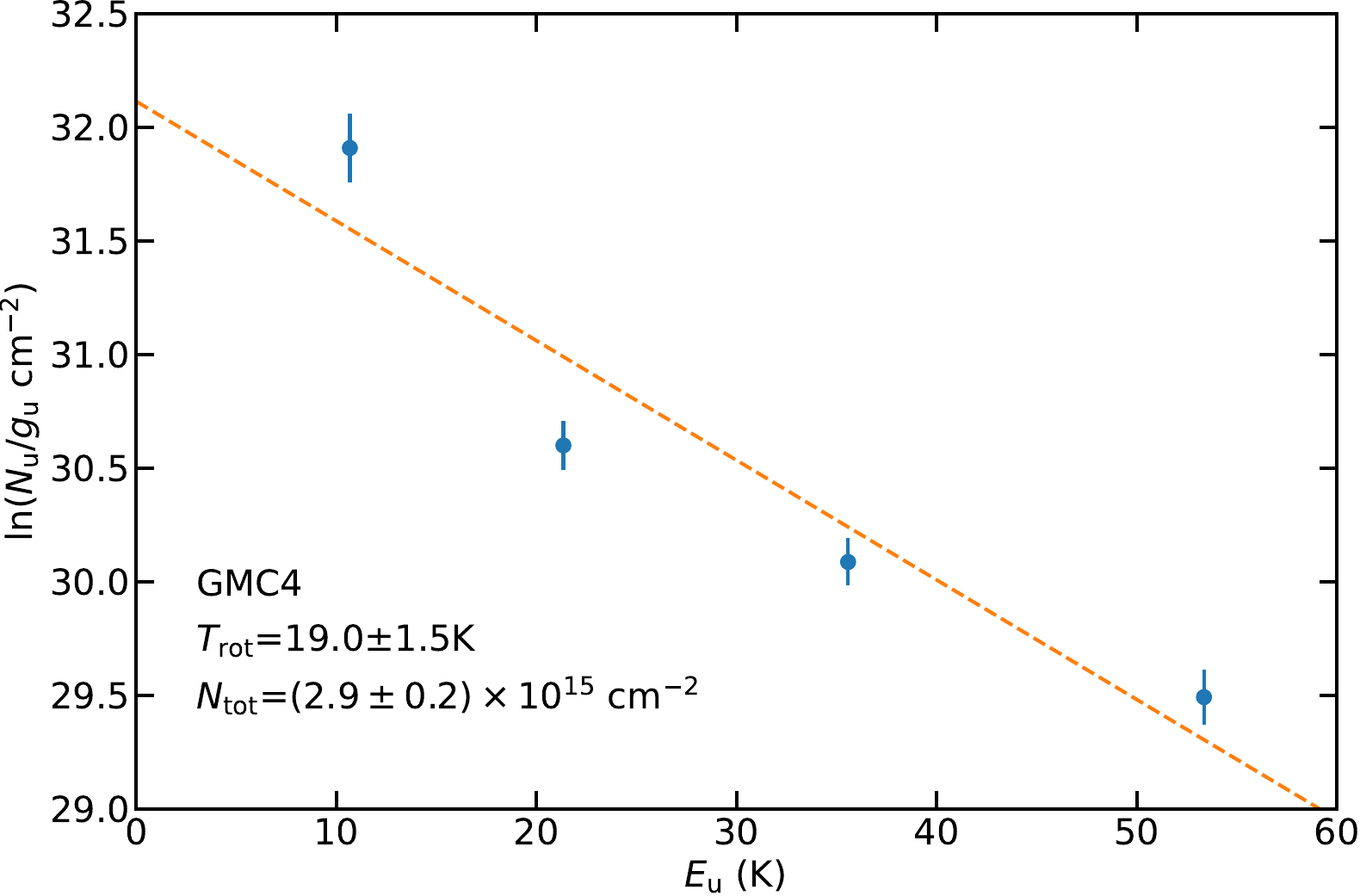}
\includegraphics[width = 0.49 \textwidth]{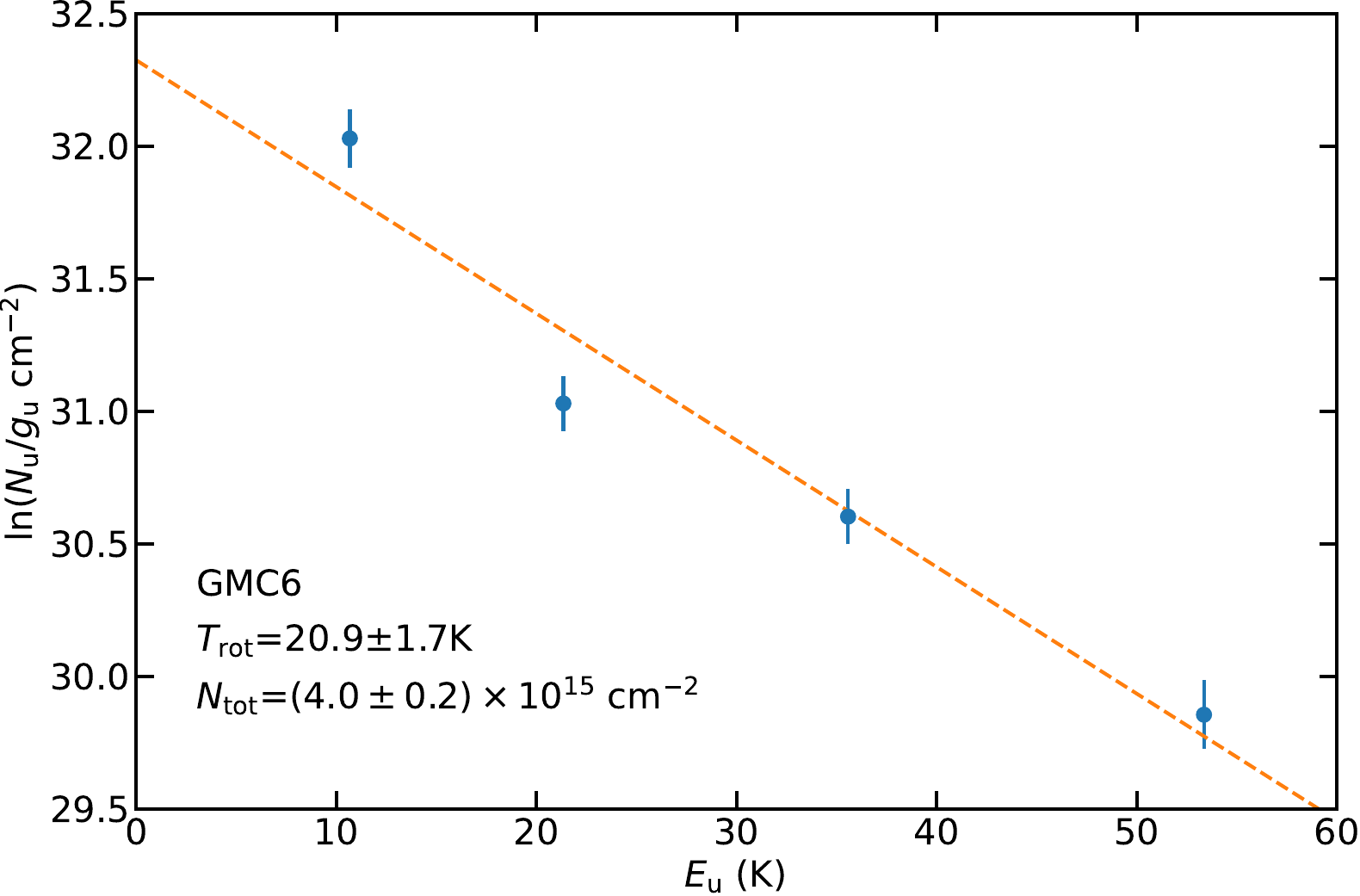}
\includegraphics[width = 0.49 \textwidth]{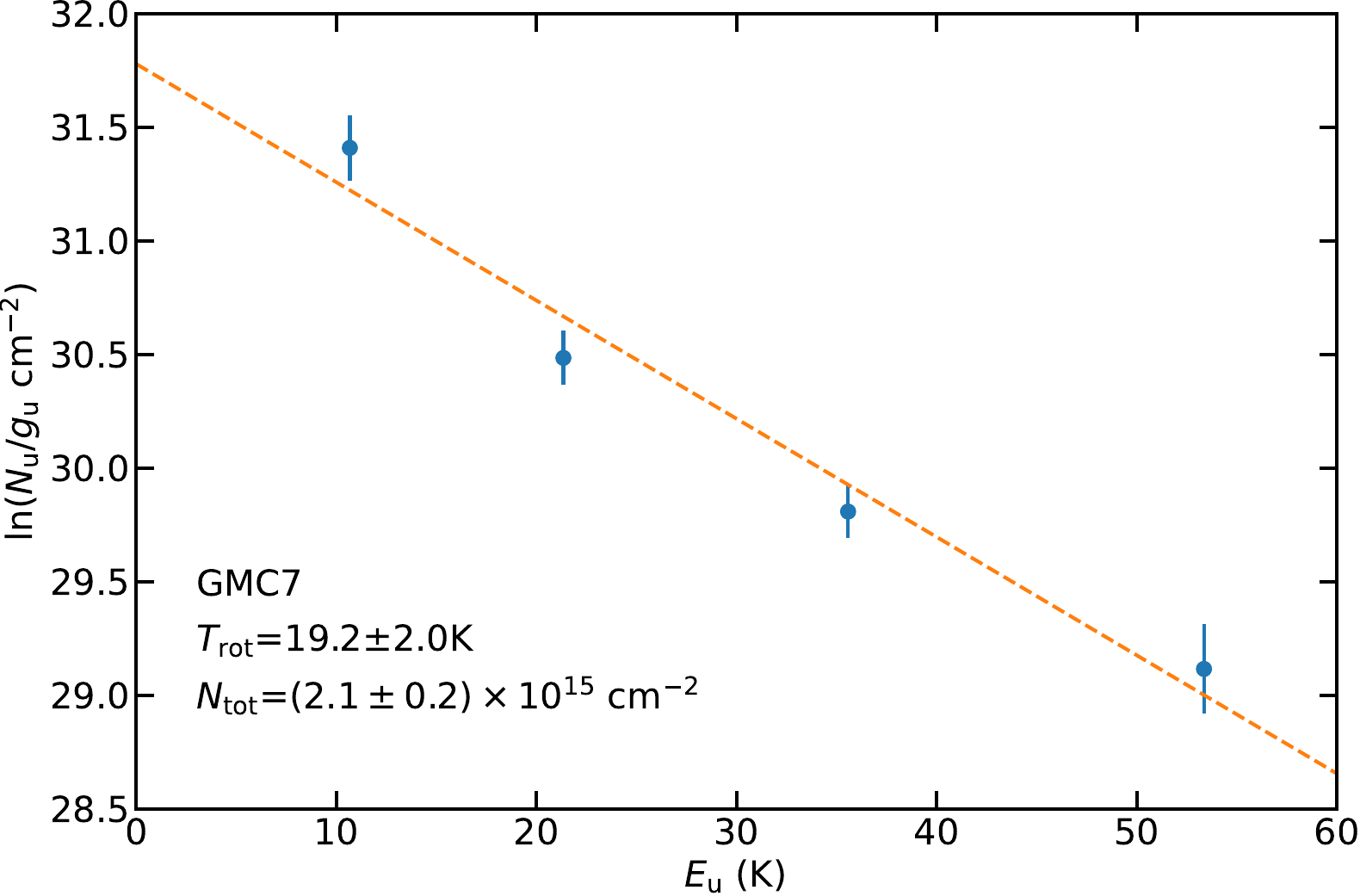}
\includegraphics[width = 0.49 \textwidth]{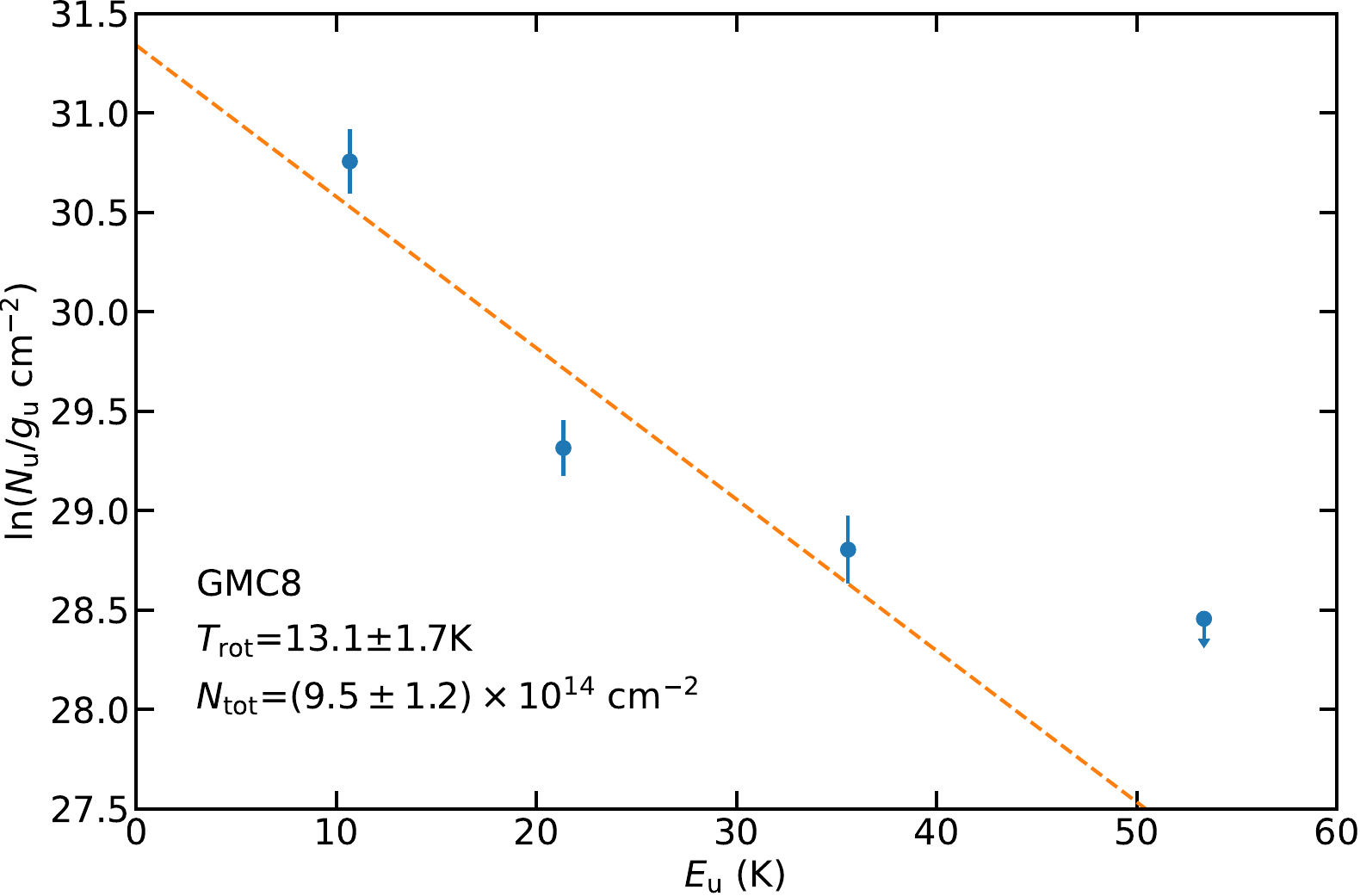}
\includegraphics[width = 0.49 \textwidth]{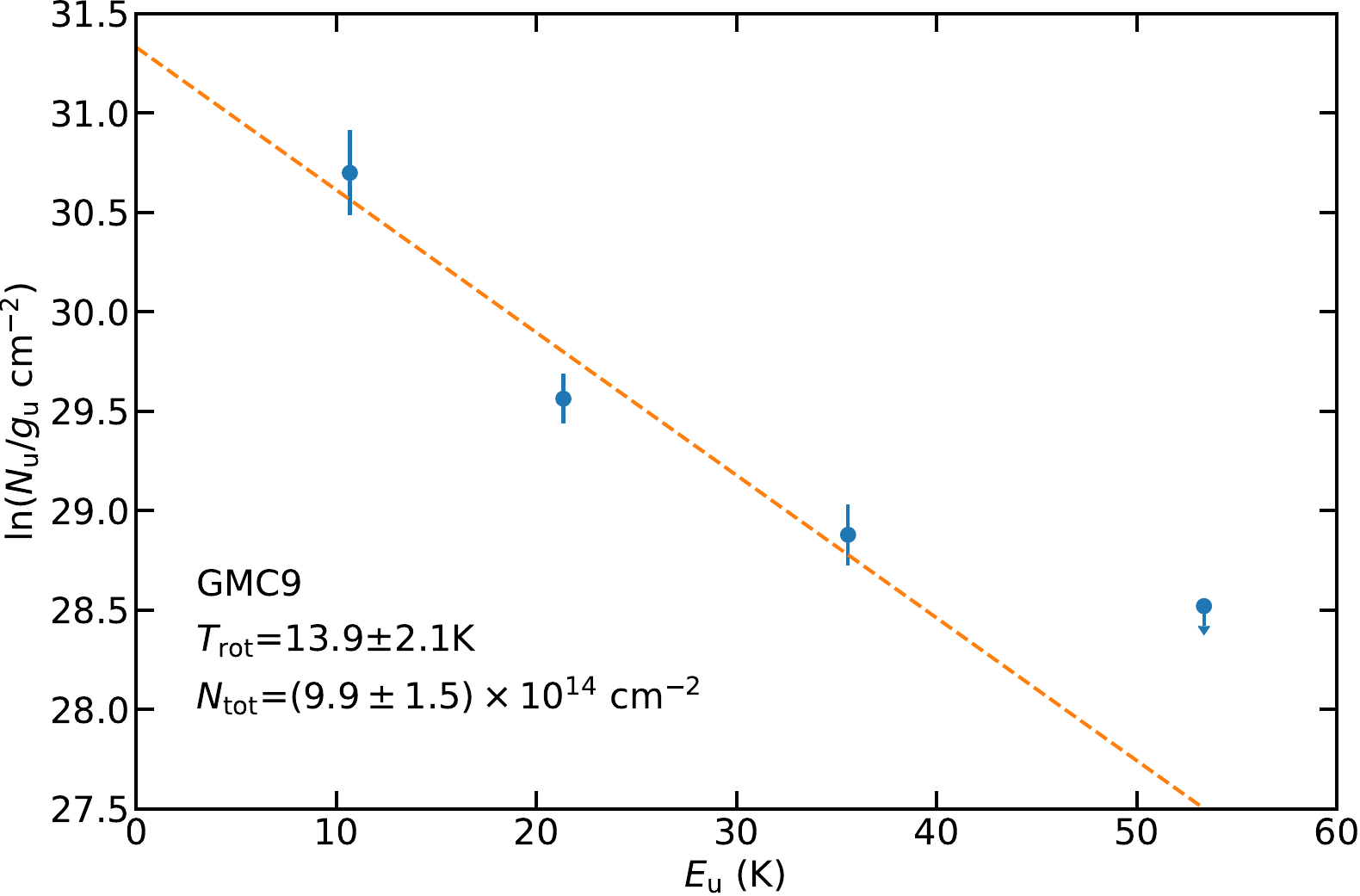}
\caption{{Rotational diagrams of HCNH$^{+}$ for the selected GMCs. In each panel, the dashed line represents the linear least squares fit to the observed data points. The fit results for each GMC are shown in the lower-left corner of each panel.}\label{Fig:rd}}
\end{figure*}

Assuming local thermodynamic equilibrium (LTE), we employed the rotational diagram method  \citep{1999ApJ...517..209G} to derive the rotational temperatures, $T_{\rm rot}$, and molecular column densities, $N_{\rm HCNH^{+}}$. Since all data have been convolved to the same angular resolution of 1\rlap{.}\arcsec6, we did not correct for the beam dilution effects, which may be negligible (see above). While this ensures that rotational temperatures are unaffected, the derived column densities are inherently beam-averaged. The rotational diagrams of the eight GMCs are shown in Fig.~\ref{Fig:rd}. Linear fits were applied to these diagrams, with uncertainties quantified using a Monte Carlo approach. Specifically, 10,000 realizations were generated by sampling Gaussian distributions of the integrated intensities, adopting the combined uncertainties from Table~\ref{Tab:obs} and 10\% of the absolute flux calibration uncertainties as the respective standard deviations. These were propagated through linear fits, and the resulting distributions were analyzed to derive uncertainties in both the rotational temperatures and column densities.

\begin{table*}[!hbt]
\caption{Physical parameters of the selected GMCs.}\label{Tab:phy}
\small
\centering
\renewcommand{\arraystretch}{1.5}
\begin{tabular}{ccccccccc}
\hline \hline
                & &\multicolumn{3}{c}{LTE} & \multicolumn{4}{c}{RADEX}                  \\
 \cmidrule(lr){3-5} \cmidrule(lr){6-9} 
Target      & $N_{\rm H_{2}}$  & $T_{\rm rot}$ &  $N_{\rm HCHN^{+}}$ & $X_{\rm HCHN^{+}}$ &  $N_{\rm HCHN^{+}}$  & $X_{\rm HCHN^{+}}$ & $T_{\rm K}$ & log$_{10}[n_{\rm H_{2}}]$   \\ 
             &  (cm$^{-2}$) & (K)           &  (cm$^{-2}$)    &                      & (cm$^{-2}$) &  & (K)  & (cm$^{-3}$)                 \\ 
(1)         & (2)        & (3)        & (4)            & (5)     & (6)          & (7)    & (8)    & (9) \\
\hline
GMC1 & $(5.0\pm1.1) \times 10^{22}$ & 17.4$\pm$2.5 & $(1.0\pm 0.1)\times 10^{15}$ & $(2.0\pm 0.5)\times 10^{-8}$  & $4.0^{+0.9}_{-0.7}\times 10^{14}$ & $8.6^{+2.4}_{-2.5}\times 10^{-9}$ &$249.83^{+101.35}_{-117.63}$ & $3.10^{+0.17}_{-0.16}$ \\
GMC2 & $(6.5\pm1.2) \times 10^{22}$ & 12.9$\pm$1.8 & $(8.3\pm 1.3)\times 10^{14}$ & $(1.3\pm 0.3)\times 10^{-8}$  & $3.2^{+1.1}_{-0.7}\times 10^{14}$ & $5.4^{+1.7}_{-1.7}\times 10^{-9}$ & $240.34^{+110.20}_{-121.93}$ & $3.01^{+0.23}_{-0.23}$ \\
GMC3 & $(2.2\pm0.5) \times 10^{23}$ & 17.7$\pm$1.4 & $(3.0\pm 0.2)\times 10^{15}$ & $(1.4\pm 0.3)\times 10^{-8}$  & $9.8^{+1.2}_{-1.0}\times 10^{14}$ & $4.8^{+1.1}_{-1.2}\times 10^{-9}$ &$278.12^{+83.78}_{-106.70}$  & $3.23^{+0.12}_{-0.11}$ \\
GMC4 & $(4.3\pm0.8) \times 10^{23}$ & 19.0$\pm$1.5 & $(2.9\pm 0.2)\times 10^{15}$ & $(6.7\pm 1.3)\times 10^{-9}$  & $7.2^{+0.8}_{-0.6}\times 10^{14}$ & $1.8^{+0.4}_{-0.4}\times 10^{-9}$ &$252.49^{+102.07}_{-132.07}$ & $3.55^{+0.22}_{-0.15}$ \\
GMC6 & $(6.3\pm1.8) \times 10^{23}$ & 20.9$\pm$1.7 & $(4.0\pm 0.2)\times 10^{15}$ & $(6.3\pm 1.8)\times 10^{-9}$  & $1.2^{+0.1}_{-0.1}\times 10^{15}$ & $2.1^{+0.6}_{-0.6}\times 10^{-9}$ &$256.86^{+99.19}_{-123.33}$  & $3.46^{+0.17}_{-0.11}$ \\
GMC7 & $(1.8\pm0.3) \times 10^{23}$ & 19.2$\pm$2.0 & $(2.1\pm 0.2)\times 10^{15}$ & $(1.2\pm 0.2)\times 10^{-8}$  & $6.7^{+0.8}_{-0.7}\times 10^{14}$ & $3.9^{+0.7}_{-0.8}\times 10^{-9}$ & $250.09^{+102.97}_{-117.32}$ & $3.35^{+0.17}_{-0.13}$ \\
GMC8 & $(5.9\pm1.1) \times 10^{22}$ & 13.1$\pm$1.7 & $(9.5\pm 1.2)\times 10^{14}$ & $(1.6\pm 0.4)\times 10^{-8}$  & $4.5^{+3.6}_{-1.5}\times 10^{14}$ & $9.4^{+4.9}_{-4.5}\times 10^{-9}$ &$254.10^{+103.72}_{-118.16}$ & $2.80^{+0.27}_{-0.33}$ \\
GMC9 & $(5.4\pm1.1) \times 10^{22}$ & 13.9$\pm$2.1 & $(9.9\pm 1.5)\times 10^{14}$ & $(1.8\pm 0.5)\times 10^{-8}$  & $3.1^{+1.1}_{-0.7} \times 10^{14}$ & $6.3^{+2.1}_{-2.1}\times 10^{-9}$ &$245.05^{+107.46}_{-119.90}$ & $3.16^{+0.25}_{-0.23}$ \\
\hline
\end{tabular}
\tablefoot{(1) Selected GMC. (2) Beam-averaged H$_{2}$ column density for a circular beam of 1\rlap{.}\arcsec6, taken from \citet{2024ApJ...961...18T}. (3) Rotational temperature. (4) Beam-averaged HCNH$^{+}$ column density, again for a circular beam of 1\rlap{.}\arcsec6. (5) HCNH$^{+}$ abundance. (6) Beam-averaged HCNH$^{+}$ column density. (7) Non-LTE HCNH$^{+}$ abundance. (8) Kinetic temperature. (9) H$_{2}$ number density. }
\normalsize
\end{table*}

The fitting results of the eight GMCs are given in Table~\ref{Tab:phy}. The derived rotational temperatures are in the range of 12--21~K, aligning closely with those of C$^{18}$O but exceeding those ($<10$~K) of H$^{13}$CO$^{+}$ \citep[see Table~4 in][]{2021ApJ...923...24H}. This discrepancy is likely attributed to the similarly low critical densities of HCNH$^{+}$ and C$^{18}$O transitions, which are much lower than those of H$^{13}$CO$^{+}$. The rotational temperatures are also lower than the expected gas temperatures \citep[e.g.,][]{2019ApJ...871..170M,2020MNRAS.491.4573R,2023A&A...675A.151H,2024ApJ...961...18T}, which implies that HCNH$^{+}$ is subthermally excited. The HCNH$^{+}$ column densities derived from our analysis range from $8.3\times 10^{14}$~cm$^{-2}$ to $4.0\times 10^{15}$~cm$^{-2}$, which is almost comparable to those of HCO$^{+}$ in this CMZ \citep{2021ApJ...923...24H}. 

%which are at least an order of magnitude higher than those obtained using the same LTE approach in the Milky Way\citep{1986ApJ...302L..31Z,1991A&A...247..487S,2021A&A...651A..94F,2023A&A...679A..39G}. This substantial increase suggests that the CMZ of NGC~253 is a large reservoir of HCNH$^{+}$ gas. 

H$_{2}$ column densities, $N_{\rm H_{2}}$, are essential to derive the fractional molecular abundances, $X_{\rm HCNH^{+}}$, of HCNH$^{+}$. In this work, we directly utilized values for $N_{\rm H_{2}}$ from the low-density hierarchical Bayesian analysis of the ALCHEMI data at a circular beam of 1\rlap{.}\arcsec6 \citep{2024ApJ...961...18T}. The resulting $X_{\rm HCNH^{+}}$, presented in Table~\ref{Tab:phy}, are remarkably high, ranging from $6.3\times 10^{-9}$ to $2.0\times 10^{-8}$. In comparison, the observed $X_{\rm HCNH^{+}}$ values in the Milky Way are $\lesssim 2\times 10^{-9}$ in all previous studies \citep[e.g.,][]{1986ApJ...302L..31Z,1991A&A...247..487S,2021A&A...651A..94F,2023A&A...679A..39G}. Particularly, $X_{\rm HCNH^{+}}$ is only $\sim 3\times 10^{-10}$ in Sgr~B2, located in the CMZ of the Milky Way \citep{1986ApJ...302L..31Z}, which is more than an order of magnitude lower than our observed values in the CMZ of NGC~253. This stark contrast highlights the significantly elevated abundance of HCNH$^{+}$ in the CMZ of NGC~253, indicating that its physical conditions differ significantly from those in the Milky Way. 
%KMM: The y-axis of Fig. 5 should be expanded, with 1x10E-10 (lower than the value for Sgr B2) as the lower bound. YG: This has been updated.

\begin{figure*}[!htbp]
\centering
\includegraphics[width = 0.95 \textwidth]{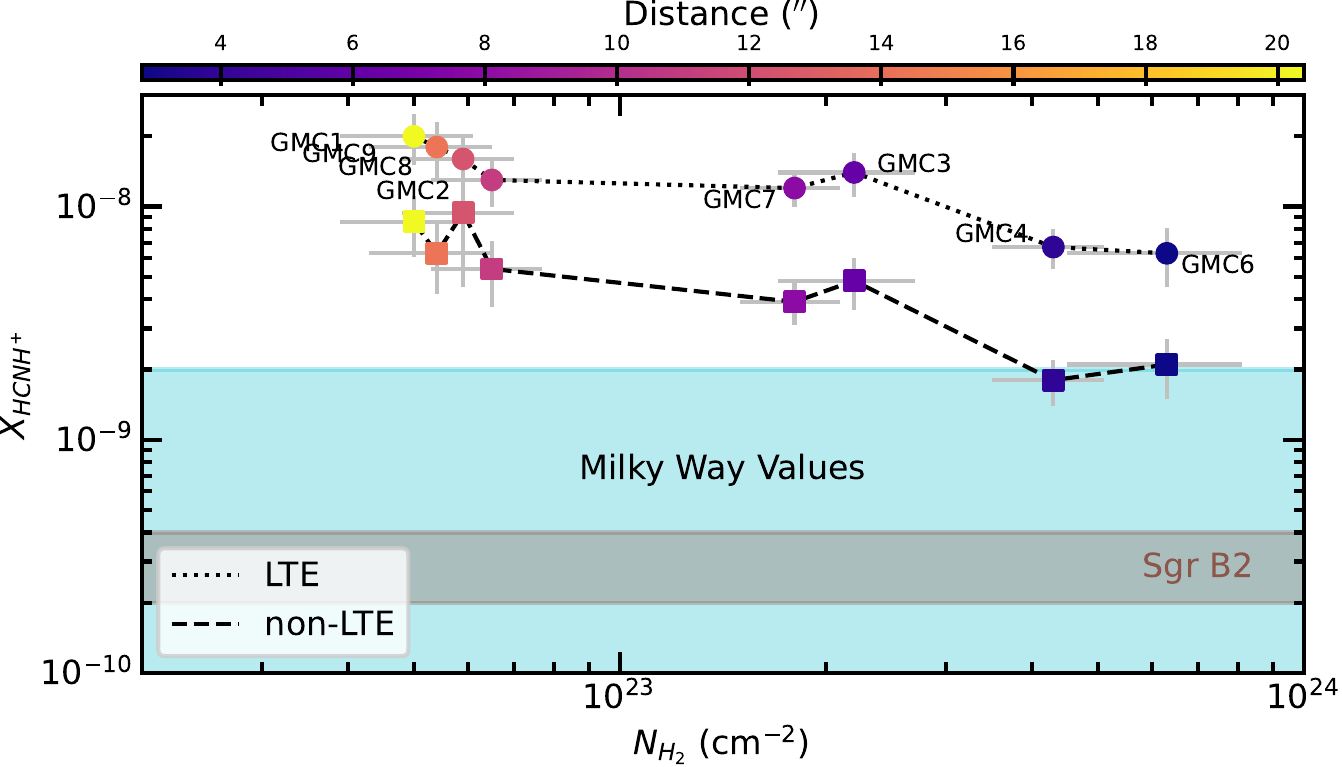}
\caption{{HCNH$^{+}$ abundances as a function of H$_{2}$ column density. The colors represent the distances of GMCs to GMC5 which is located at the center of the CMZ of NGC~253. The circles represent HCNH$^{+}$ abundances derived using the LTE approach (see Sect.~\ref{Sec:res}), while the squares represent those obtained from the non-LTE approach (see Sect.~\ref{sec.density}). The cyan-shaded region indicates the observed HCNH$^{+}$ abundances of $\lesssim 2\times 10^{-9}$ previously reported in the Milky Way, while the brown-shaded region marks the abundances reported toward Sgr~B2.}\label{Fig:anti}}
\end{figure*}

Based on Table~\ref{Tab:phy}, we observe a trend of decreasing HCNH$^{+}$ abundances in GMCs as they approach the center of the CMZ, which is similar to the gradient of HOCO$^{+}$ abundances \citep[see Fig.~4d in][]{2022ApJ...938...80H}. To further explore this, we examined the relationship between HCNH$^{+}$ abundance and H$_2$ column density, as shown in Fig.~\ref{Fig:anti}. Our analysis reveals that the anti-correlation between HCNH$^{+}$ abundances and H$_2$ column densities, previously observed in molecular clouds of the Solar neighborhood \citep{2023A&A...679A..39G}, also holds in the CMZ of NGC 253. However, the anti-correlation appears to arise from the associated cosmic ray ionization rates (see Sect.~\ref{Sec:dis}), because GMCs close to the center of the CMZ tend to have higher cosmic ray ionization rates \citep{2021ApJ...923...24H,2022ApJ...931...89H,2022ApJ...939..119B,2024ApJ...977...38B}. In the Solar neighborhood, cosmic ray ionization rates are likely governed by H$_2$ column densities. As suggested by previous studies \citep[e.g.,][]{2017ApJ...845..163N,2018A&A...614A.111P}, cosmic ray ionization rates decrease with increasing H$_2$ column density. If these rates were the primary factor regulating HCNH$^{+}$ abundances, one would expect higher HCNH$^{+}$ abundances in regions with higher H$_2$ column densities. However, this expectation is not consistent with recent observations. Therefore, we propose that the physical mechanisms driving the observed anti-correlation between HCNH$^{+}$ abundance and H$_2$ column density in the CMZ of NGC 253 differ from those in the Solar neighborhood.

%Despite the starkly different environments, our findings confirm that HCNH$^{+}$ tends to be depleted in regions with high H$_2$ column densities. This depletion might be indicative of chemical processes that preferentially destroy HCNH$^{+}$ in denser regions, regardless of the galactic environment.

%\subsection{Comparison with its precusor molecules}
%correlation with HCN/HNC.
%We demonstrate the HCNH$^{+}$ emission is more widespread than the one presented in the previous work. 
%Hence, we used the fitted velocity centroids and line widths derived from HCNH$^{+}$ (2--1) as the input parameters to fit the high $J$ transitions of HCNH$^{+}$.

%HCNH$^{+}$ prevalent in the CMZ of NGC~253. 
\subsection{Non-LTE analysis}
\subsubsection{Collisional rate coefficients}\label{sec.coll}
Rates for collisions of HCNH$^{+}$ induced by encounters with {\it para}-H$_2$($j_2$\footnote{$j_2$ stands for the rotational quantum number of H$_2$.} = 0), denoted hereafter as {\it p}-H$_2$, were first investigated by \cite{2023JChPh.158g4304B} using the time-independent quantum mechanical close-coupling (CC) method \citep{arthurs1960theory} and later revised to include the effect of the H$_2$ rotational basis \citep{2023A&A...679A..39G,2024A&A...681L..19B}. The former dataset, which is less precise, covers kinetic temperatures, $T_{\rm K}$, up to 100~K, while the latter more precise dataset is limited to 30~K due to the computational cost of the CC method. Given that the CMZ of NGC~253 is expected to be significantly warmer than 30~K \citep[e.g.,][]{2019ApJ...871..170M,2024ApJ...961...18T}, a comprehensive set of collisional rate coefficients covering a wider temperature range is required for a robust interpretation of warm or hot HCNH$^{+}$ emission lines in the CMZ. This would help in accurately modeling the excitation conditions of HCNH$^{+}$ in such extreme environments.

To address the computational cost of the CC method, we employ the quantum mechanical coupled states (CS) approximation \citep{mcguire1974quantum}, which has long been tested and validated for the computation of high-temperature collisional rate coefficients. We investigate the scattering of HCNH$^+$ due to collisions with both {\it p}-H$_2$ and {\it ortho}-H$_2$($j_2$ = 1), denoted hereafter as {\it o}-H$_2$. The use of the latter collision partner is motivated by the warm/hot regime which we are interested in. Indeed, above 75~K, the population of molecular hydrogen is dominated by the {\it o}-H$_2$ spin-symmetry.

\begin{figure}
    \centering
    \includegraphics[width = 0.98\linewidth, trim = 0 10 10 10, clip = true]{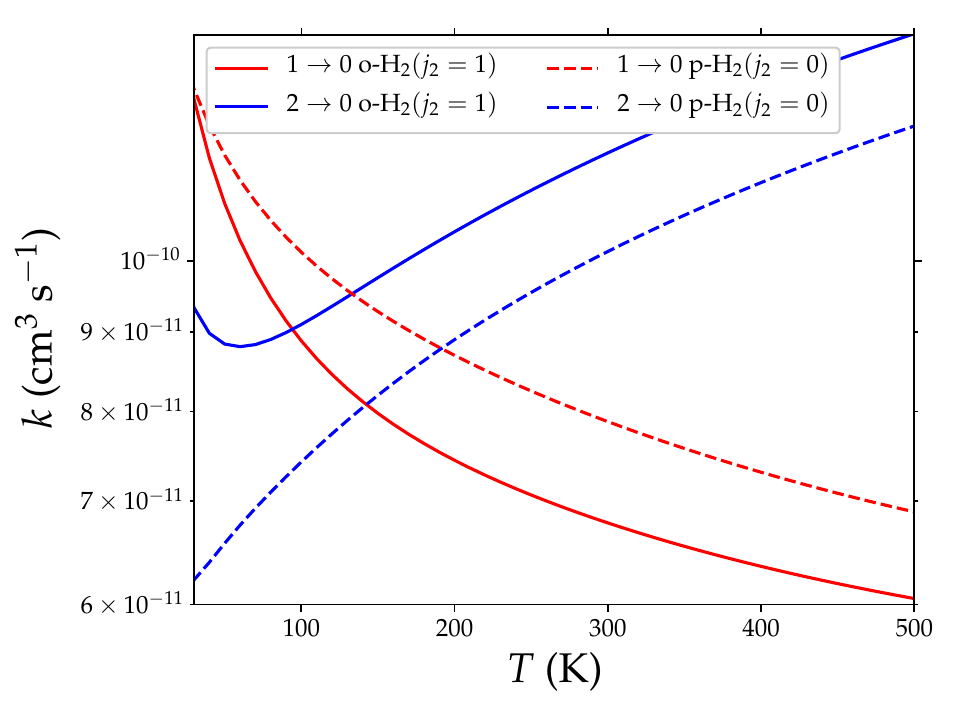}
    \caption{Temperature dependence of the HCNH$^+$ rate coefficients induced by collisions with {\it p}-H$_2$ (dashed lines) and {\it o}-H$_2$ (solid lines).}
    \label{fig:rates}
\end{figure}

In practice, we implemented the four-dimensional potential energy surface calculated by \cite{2023JChPh.158g4304B} in the \texttt{MOLSCAT} molecular scattering code \citep{hutson1994molscat} to compute state-to-state inelastic cross sections. We focus on transitions among the 30 lowest-lying energy levels of HCNH$^+$, which, for convergence reasons, require an exhaustive rotational basis ($j_1$\footnote{$j_1$ stands for the rotational quantum number of HCNH$^+$.} = $0-48$). In addition, we include the two low-lying rotational transitions of H$_2$, namely, $j_2=0-2$ and $j_2=1-3$ for {\it p}- and {\it o}-H$_2$, respectively. All additional integration parameters are defined to ensure subpercent convergence of the cross sections. The calculations were performed for total energies up to 3500~cm$^{-1}$ and 3620~cm$^{-1}$ using {\it p}-H$_2$ and {\it o}-H$_2$ as colliders, respectively. Therefore, using the Maxwell-Boltzmann velocity distribution, we derive rate coefficients for HCNH$^+$ induced by collisions with both {\it p}- and {\it o}-H$_2$ for kinetic temperatures up to 500~K, and the results are shown in Fig.~\ref{fig:rates}. The collisional rate coefficients will be accessible through databases such
as Basecol\footnote{\url{https://basecol.vamdc.eu/}} \citep{2024A&A...683A..40D}, EMAA\footnote{\url{https://emaa.osug.fr/}}, and LAMDA\footnote{\url{https://home.strw.leidenuniv.nl/~moldata/}} \citep{2005A&A...432..369S,2020Atoms...8...15V}.

\begin{figure}
    \centering
    \includegraphics[width = 0.98\linewidth, trim = 10 10 10 10, clip = true]{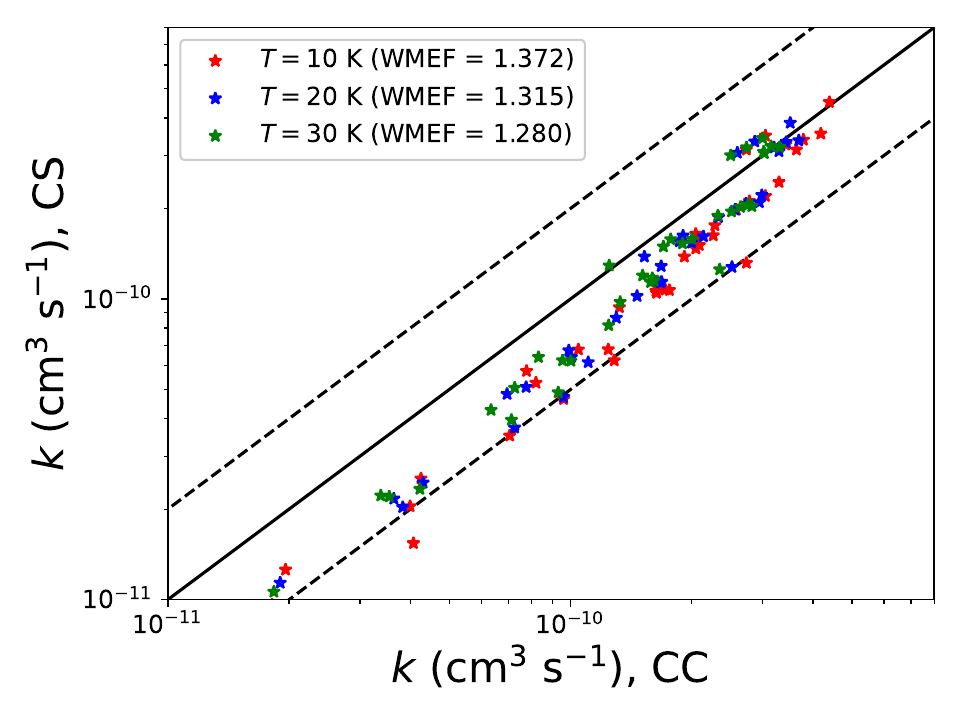}
    \caption{Comparison of two datasets of rate coefficients for HCNH$^+$ induced by collisions with {\it p}-H$_2(j_2=0)$. The x-axis represents the first dataset, computed using the ``exact" CC method \citep{2023JChPh.158g4304B}, while the y-axis represents the second dataset, calculated in this work using the CS approximation. The dashed lines indicate an agreement region within a factor of 2.0. Different colors represent their weighted mean error factors (WMEF) which are defined in Eq.~(\ref{f.wmef}).}
    \label{fig:c_vs_cs}
\end{figure}

To assess the accuracy of the CS approximation, we compare in Fig.~\ref{fig:c_vs_cs} the new HCNH$^+$ collisional rate coefficients induced by {\it p}-H$_2$ with those previously computed by \cite{2024A&A...681L..19B} using the ``exact" CC method. The collisional rate coefficients using the CS approximation turn out to underestimate the CC data by up to a factor of two, especially for the low-magnitude rate coefficients. To quantify the disagreement between the CC and the CS results, we calculate for each temperature the weighted mean error factor \citep{loreau2018efficient,pirlot2023hyperfine} as follows:

\begin{eqnarray}\label{f.wmef}
    {\rm WMEF} = \frac{\sum_i k_i^{\rm CC}\times r_i}{\sum_i k_i^{\rm CC}},
\end{eqnarray}
where $r_i$ is defined as the maximum of $k_i^{\rm CC}/k_i^{\rm CS}$ or $k_i^{\rm CS}/k_i^{\rm CC}$, ensuring that $r_i\geq1$. The WMEF decreases with increasing temperature, suggesting that the agreement is expected to be reasonable above 30~K, good for $T > 50$~K, and excellent above 80~K. For further validation tests, we refer readers to Appendix~\ref{appendix_Tr_ratio}, where the impact of low-temperature discrepancies on radiative transfer calculations is shown.

\subsubsection{Non-LTE modeling}\label{sec.density}
The detection of multiple HCNH$^{+}$ transitions in the CMZ of NGC~253 allows the first non-LTE analysis of HCNH$^{+}$ in an extragalactic environment. Here, we used the non-LTE radiative transfer code RADEX\footnote{\url{https://home.strw.leidenuniv.nl/~moldata/radex.html}} for statistical equilibrium calculations \citep{2007A&A...468..627V}. The collisional rates used in the analysis are based on the data discussed in Sect.~\ref{sec.coll}. In our modeling, RADEX adopts an expanding spherical geometry, the so-called Sobolev or large velocity gradient (LVG) approximation \citep[e.g.,][]{1960mes..book.....S,1992ARA&A..30...75E}, to calculate the escape probability. Given that HCNH$^{+}$ transitions are optically thin in the CMZ of NGC 253 (see Sect.~\ref{Sec:res}), the results are not affected by the assumed geometry. Thus, an expanding spherical geometry providing localized source functions is sufficiently robust.

In our RADEX modeling, we adopted an H$_{2}$ ortho-to-para ratio of 0.25, following the values reported in previous studies \citep[e.g.,][]{2006ApJ...649..816N}. The ortho-to-para ratio also does not affect the modeling results significantly, because the rate coefficients for collisions with {\it p}- and {\it o}-H$_2$ are comparable (see Fig.~\ref{fig:ortho_para-T_50-500} in Appendix~\ref{appendix_rates}). The line widths are taken to be the averages of the fitted values for the four transitions  (see Table~\ref{Tab:obs}). We assume that the emission from all HCNH$^{+}$ transitions originates from the same regions. Given that the beam sizes are identical for all transitions, beam dilution effects primarily influence the modeling of HCNH$^{+}$ column densities when the transitions are optically thin. For simplicity, we neglected the beam dilution effects (i.e., the beam dilution factor is set to be unity). As a result, the HCNH$^{+}$ column densities derived from our non-LTE analysis are beam-averaged values. 

We minimize our modeling results using the emcee\footnote{\url{https://emcee.readthedocs.io/en/stable/}} code \citep{2013PASP..125..306F} and the affine-invariant ensemble sampler \citep{2010CAMCS...5...65G} for Monte Carlo
Markov chain (MCMC) calculations, enabling us to obtain the posterior probability distributions of the modeled parameters. We assumed uniform priors for $T_{\rm K}$, $n_{\rm H_2}$, and $N_{\rm HCNH^{+}}$. Based on previous studies on the CMZ of NGC~253 \citep[e.g.,][]{2019ApJ...871..170M,2022ApJ...931...89H,2022ApJ...939..119B,2024ApJ...961...18T}, we set $T_{\rm K}$, $n_{\rm H_2}$, and $N_{\rm HCNH^{+}}$ to be in the range of 50--500~K, 10$^{2}$--10$^{7}$~cm$^{-3}$, and 10$^{13}$--10$^{17}$~cm$^{-2}$, respectively. The posterior distribution of these parameters is determined by the product of the prior distribution function and the likelihood function.
The likelihood function is assumed to be e$^{-\chi^{2}/2}$ with 
\begin{equation}
    \chi^{2} = \Sigma_{i}(I_{{\rm obs},i}-I_{{\rm mod},i})/\sigma_{i} \;,
\end{equation}
where $I_{{\rm obs},i}$, $I_{{\rm mod},i}$, and $\sigma_{i}$ are the observed integrated intensity ratio, the modeled integrated intensity, and the 1$\sigma$ uncertainty in $I_{{\rm obs},i}$, respectively. The MCMC simulations were conducted using 20 walkers for 4500 steps following a burn-in period to ensure that the walkers achieved convergence. The 1$\sigma$ uncertainties of the physical parameters are derived from the 16th and 84th percentiles of the posterior distribution.  

\begin{figure*}[!htbp]
\centering
\includegraphics[width = 0.95 \textwidth]{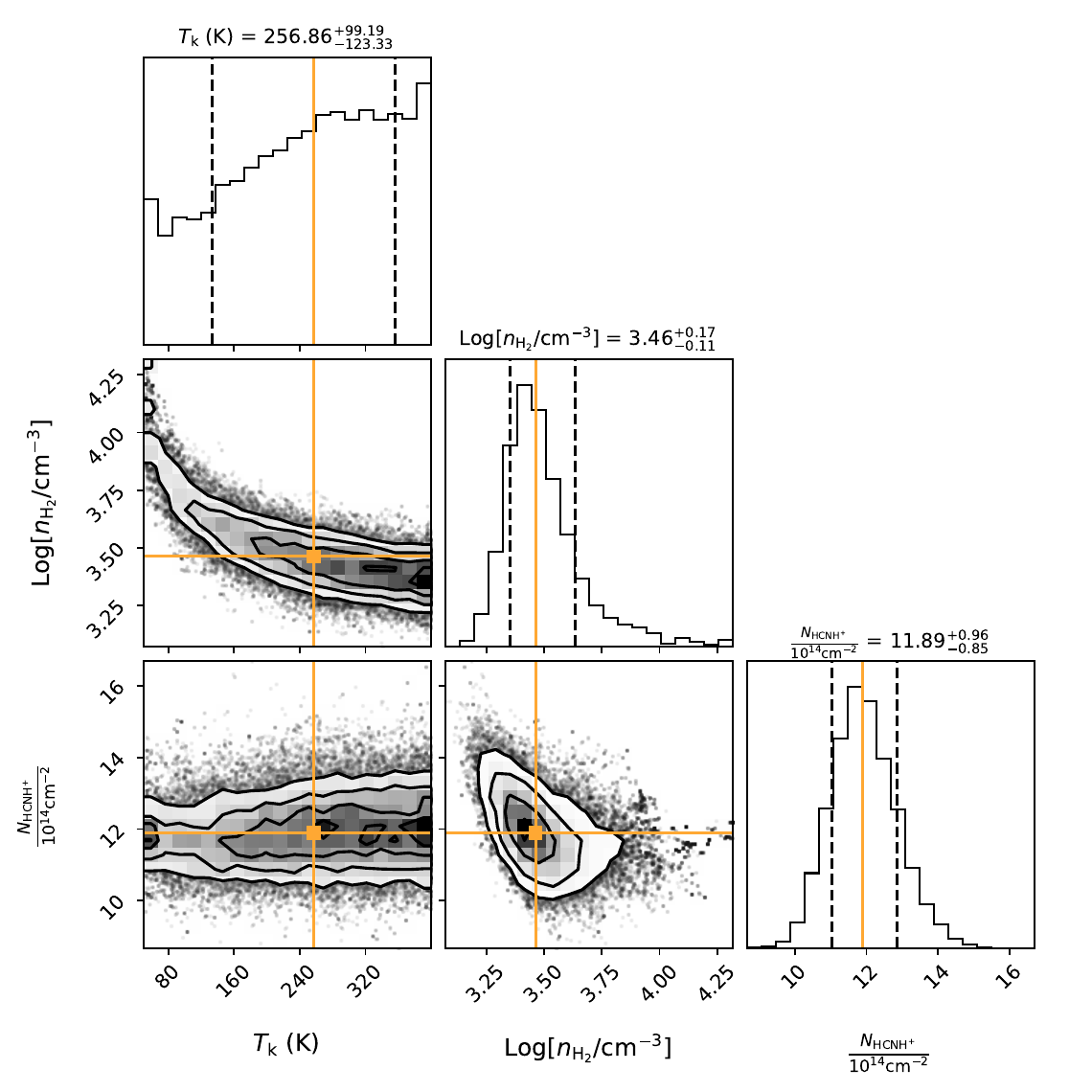}
\caption{{Posterior probability distributions of gas temperatures, H$_{2}$ number densities, and HCNH$^{+}$ column densities for GMC6, with the maximum posterior possibility point in the parameter space highlighted by orange lines and points. Contours represent the 0.5, 1.0, 1.5, and 2.0$\sigma$ confidence intervals. The vertical dashed lines represent the 1$\sigma$ spread.}\label{Fig:mcmc}}
\end{figure*}

Figure~\ref{Fig:mcmc} shows an example of the fitting results for GMC6. The plot demonstrates that H$_{2}$ and HCNH$^{+}$ column densities are well constrained, whereas the gas temperatures are less precisely determined. The large uncertainties associated with the gas temperatures suggest that HCNH$^{+}$ transitions are relatively insensitive to variations in gas temperature within the parameter space considered in the models. The complete fitting results for the eight GMCs are summarized in Table~\ref{Tab:phy}. 

The beam-averaged HCNH$^+$ column densities of $(0.3-1.2) \times 10^{15}$~cm$^{-2}$, derived from the best-fit RADEX modeling, are about a factor of three lower than those obtained from the LTE approach. This discrepancy suggests that the HCNH$^+$ gas deviates from LTE conditions, despite its low critical densities. Such behavior indicates that HCNH$^+$ predominantly arises from low-density regions, where H$_2$ number densities are below $10^4$~cm$^{-3}$ \citep{2024A&A...681L..19B}. Therefore, we consider the HCNH$^+$ column densities and abundances derived from our RADEX modeling to be more reliable. Even with the low HCNH$^+$ column densities derived from non-LTE modeling, the HCNH$^+$ abundances are revised to (1.8--9.4)$\times 10^{-9}$ in the CMZ of NGC~253, most of which remain higher than all reported Galactic HCNH$^{+}$ abundances ($\lesssim 2\times 10^{-9}$) \citep[e.g.,][]{1986ApJ...302L..31Z,1991A&A...247..487S,2021A&A...651A..94F,2023A&A...679A..39G}.

Given the low dipole moment of HCNH$^{+}$ \citep[0.29~D;][]{1986CPL...124..382B}, the critical densities of its transitions are relatively low \citep[see Table 1 in][]{2023A&A...679A..39G}. Utilizing the new collisional rate coefficients, we calculated the H$_{2}$ critical densities of the four HCNH$^{+}$ transitions at high kinetic temperatures using Eq.~(5) in \citet{2015PASP..127..299S}. At $T_{\rm k}=100$~K, the critical densities for HCNH$^{+}$ (2--1), HCNH$^{+}$ (3--2), HCNH$^{+}$ (4--3), HCNH$^{+}$ (5--4) in the optically thin case are $1.0\times 10^{3}$~cm$^{-3}$, $3.6\times 10^{3}$~cm$^{-3}$, $8.6\times 10^{3}$~cm$^{-3}$, and $1.7\times 10^{4}$~cm$^{-3}$, respectively, which are much lower than those of the transitions of other molecules like HCN, HNC, CN, HCO$^{+}$, and N$_{2}$H$^{+}$ with comparable $J$ quantum numbers \citep[see Table~1 in][]{2015PASP..127..299S}. When $T_{\rm k}>100$~K, the critical densities decrease slightly. These critical densities indicate that they trace low-density gas with $n_{\rm H_{2}}\lesssim 10^{4}$~cm$^{-3}$. Our non-LTE modeling analysis indeed reveals that the derived H$_{2}$ number densities are $\sim 10^{2.80}-10^{3.55}$~cm$^{-3}$ (see Table~\ref{Tab:phy}), aligning with the values of the low-density component in the CMZ of NGC~253 \citep[e.g.,][]{2024ApJ...961...18T}. Furthermore, the H$_{2}$ number densities of the inner GMCs (i.e., GMC4 and GMC6) exceed those in the outer regions, consistent with the trend of the low-density component characterized in Fig.~5 of \citet{2024ApJ...961...18T}. This agreement confirms that HCNH$^{+}$ primarily traces the low-density regions of GMCs, highlighting its potential as a reliable tracer of such environments.%, even in extreme environments. 

% one surprising result is that HCN and HCNH+ trace different density regimes. Owing to their high dipole moments, \citet{2022ApJ...939..119B,2024ApJ...961...18T} modeled HCN and HNC, and found their H2 number density about 1e5.5 cm$^{-3}$  which is nearly an order of magnitude higher than the density regime traced by HCNH$^{+}$. The different excitation coditions indicates that they might not be tracing the same gas. 

%HCN and HCNH$^{+}$ are thought to be a chemically coevolving pair \citep[e.g.,][]{1991A&A...247..487S,2023A&A...679A..39G,2023MNRAS.521.5087T}. In the CMZ of NGC~253, HCN is thought to trace the high-density component \citep{2024ApJ...961...18T}.

%Due to the high dipole moment, HCN commonly traces dense gas, different from HCNH$^{+}$.  

%which was proposed to study ambipolar diffusion in light of simulations (Tritsis). However, our modeling show  they trace different density regmie. Therefore, we conclude that this cheimical pair is supposed to be not good for studying ambipolar diffusion because of their different excitation conditions. 
\section{Discussion}\label{Sec:dis}
%\subsection{Chemical enhancement of HCNH$^{+}$ abundances}\label{sec.chem}
In this study, we derived surprisingly high HCNH$^{+}$ abundances in the CMZ of NGC~253, exceeding all reported values in the Milky Way. This stark contrast suggests that HCNH$^{+}$ can be more abundant than previously thought, which underscores the environmental influences on its abundance. Compared to previous studies in the Milky Way, the extreme physical conditions in the CMZ of NGC~253 might potentially enhance the formation of HCNH$^{+}$. Therefore, we performed astrochemical calculations to study the environmental influences on HCNH$^{+}$ abundances using models of static clouds, photodissociation regions (PDRs), and shocks.

\subsection{Static clouds}\label{sec.mod1}
We first used the gas-grain astrochemical code, \textit{Chempl\footnote{\url{https://github.com/fjdu/chempl}}} \citep{2021RAA....21...77D}, for calculations of a static cloud, employing the UMIST RATE12 chemical network \citep{2013A&A...550A..36M} enhanced with updated chemical reactions as described by \citet{2023A&A...679A..39G}. The initial elemental abundances are the same as in Table~3 of \citet{2013A&A...550A..36M}, who tested the network using a dark cloud model. Although 
their initial conditions may not precisely match those of the CMZ of NGC~253, the environmental influences on HCNH$^{+}$ abundances are similar for different initial conditions. Therefore, our modeling results can provide an illustrative case for understanding the environmental influences on the trends in abundance variations.

Our investigation focused on the effects of various parameters, including gas temperatures, gas densities, visual extinctions, interstellar radiation, and cosmic ray ionization rates on HCNH$^{+}$ abundances. In these models, dust temperatures are assumed to be equal to gas temperatures.
%%% KMM: Is this realistic? In the MW's CMZ, T_dust is significantly lower than T_kin. Is this also so in NGC 253's CMZ?
Based on previous studies, we constrained gas temperatures to $T_{\rm K}\lesssim$500~K \citep{2019ApJ...871..170M,2020MNRAS.491.4573R,2023A&A...675A.151H,2024ApJ...961...18T}, gas densities to $n_{\rm gas}=n_{\rm H}+n_{\rm H_{2}} \lesssim 10^{5}$~cm$^{-3}$ (see Sect.~\ref{sec.density} and \citealt{2024ApJ...961...18T}), interstellar radiation fields to $G_{0}\lesssim 10^{9}$ in Habing units \citep{2018ApJ...860...23P}, and cosmic-ray ionization rates to $\zeta \lesssim 10^{5}\zeta_{0}$ \citep[e.g.,][]{2021ApJ...923...24H,2022ApJ...931...89H,2022ApJ...939..119B,2024ApJ...977...38B} with $\zeta_{0}=1.36\times 10^{-17}$~s$^{-1}$ \citep[e.g.,][]{2000A&A...358L..79V}.

\begin{figure*}[!htbp]
\centering
\includegraphics[width = 0.95 \textwidth]{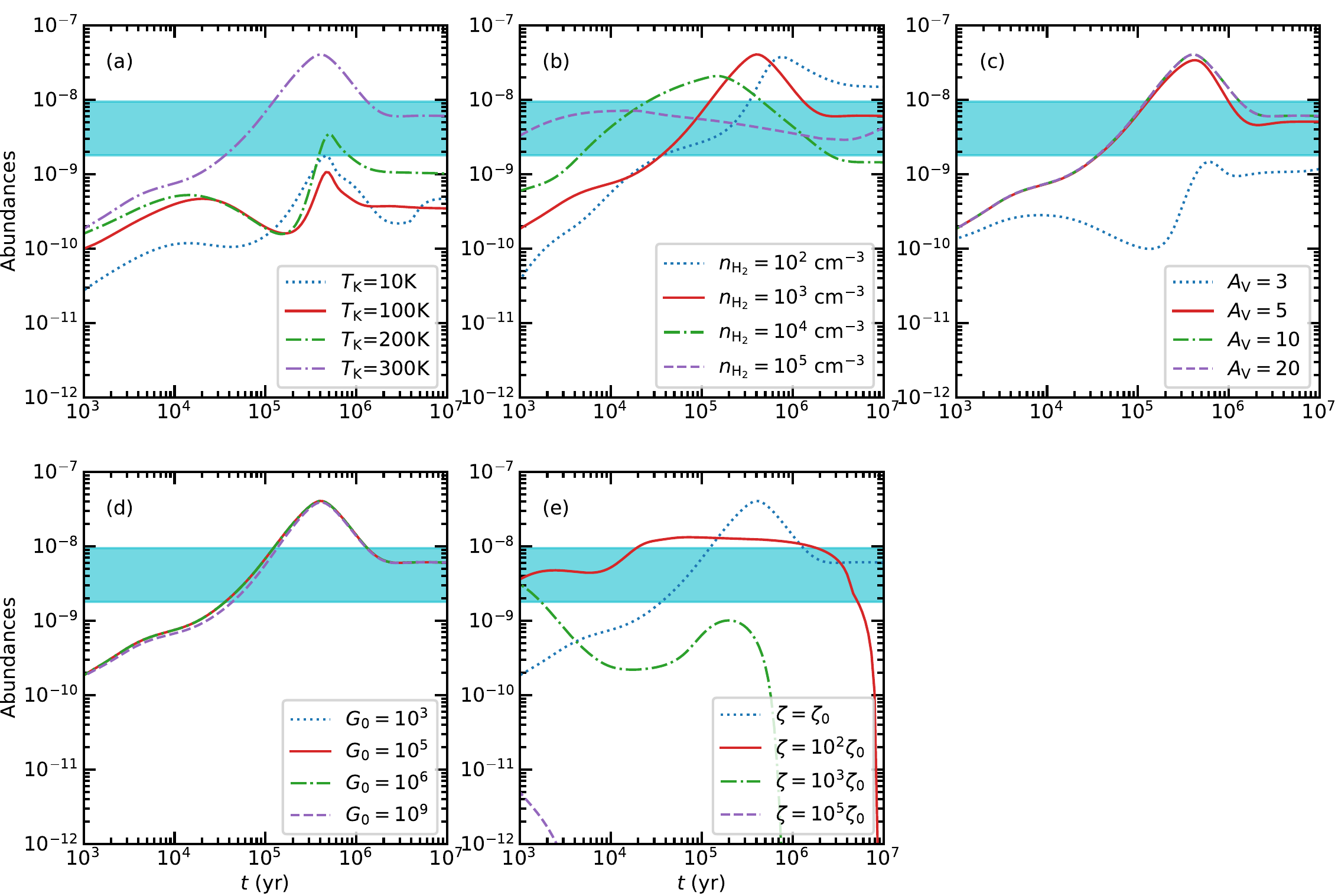}
\caption{{Environmental influence on HCNH$^{+}$ abundances relative to H$_{2}$ as a function of time, calculated from the static models of \textit{Chempl} \citep{2021RAA....21...77D}. (a) Modeling results for different gas temperatures with $n_{{\rm H}_{2}}=10^{3}$~cm$^{-3}$, $A_{\rm v}=$10, $G_{0}$=1, and $\zeta=\zeta_{0}$. (b) Modeling results for different H$_{2}$ number densities with $T_{\rm K}=300$~K, $A_{\rm v}=$10, $G_{0}$=1, and $\zeta=\zeta_{0}$. (c) Modeling results for different visual extinctions with $T_{\rm K}=300$~K, $n_{{\rm H}_{2}}=10^{3}$~cm$^{-3}$, $G_{0}$=1, and $\zeta=\zeta_{0}$. (d) Modeling results for different interstellar radiation fields with $T_{\rm K}=300$~K, $n_{{\rm H}_{2}}=10^{3}$~cm$^{-3}$, $A_{\rm v}=$20, and $\zeta=\zeta_{0}$. (e) Modeling results for different cosmic ray ionization rates with $T_{\rm K}=300$~K, $n_{{\rm H}_{2}}=10^{3}$~cm$^{-3}$, $A_{\rm v}=$20, and $G_{0}=10^{5}$. In all panels, the cyan-shaded regions indicate the observed HCNH$^{+}$ abundances derived from our non-LTE analysis (see Table~\ref{Tab:phy}).}\label{Fig:chempl}}
\end{figure*}

Figure~\ref{Fig:chempl} illustrates the modeling results under different physical conditions. In Fig.~\ref{Fig:chempl}a, HCNH$^{+}$ abundances at $T_{\rm K}=100$~K can be even lower than those at $T_{\rm K}=10$~K at $10^{6}$~yr, but HCNH$^{+}$ abundances increase with increasing gas temperatures above $T_{\rm K}=100$~K at $\gtrsim 10^{6}$~yr. This transition is related to the evaporation temperatures, $T_{\rm evap}$, of HCNH$^{+}$'s precursor molecules such as HCN, above which HCN can be effectively released from dust grains. The evaporation temperature, at which the concerned molecule is half in gas phase and half on grain surfaces, can be estimated with the following formula \citep[e.g.,][]{2009ApJ...690.1497H,2012A&A...538A..91D}: 
\begin{equation}
    T_{\rm evap} \simeq E_{\rm D}/\left\{ 60+ {\rm ln} \left[ \left(\frac{10^{5}\;{\rm cm^{-3}}}{n_{\rm H}}\right ) \left (\frac{\rm 20\;K}{T_{\rm gas}}\right )^{1/2} \left (\frac{m}{\rm 28\;amu}\right )^{1/2}\right] \right\}
\end{equation}
where $E_{\rm D}$ is the binding energy barrier of the species on grain surfaces, $n_{\rm H}$ is the hydrogen number density, $T_{\rm gas}$ is the gas temperature, and $m$ is the molecular mass of the species. HCN binding energies were estimated to vary within the range of 2057--7421~K, depending on the different surface models \citep{2020ApJ...904...11F}. Based on the results from experiments \citep{2022ApJ...933..206B}, we adopted HCN–HCN and HCN-H$_{2}$O binding energies of $E_{\rm D}=3207\pm 197$~K and $E_{\rm D}=4192\pm 68$~K for our calculations. For HCN, $m=$27~amu. Therefore, we estimate the evaporation temperatures of about 50--70~K for HCN molecules. In the CMZ of NGC~253, the gas temperature ranges from $\lesssim$50K on larger scales to $\gtrsim$300~K on 10~pc scales \citep{2019ApJ...871..170M}. On a scale of 27~pc (i.e., our beam size of 1\rlap{.}\arcsec6), the gas temperatures are likely higher than the HCN evaporation temperatures \citep{2019ApJ...871..170M,2023A&A...675A.151H,2024ApJ...961...18T,2024A&A...689A..64B}. These elevated gas temperatures can significantly enhance the formation of the gas-phase precursors of HCNH$^+$ (i.e., HCN and HCN$^{+}$), thereby increasing its abundance. 
%Notably, our model at a fiducial temperature of $T_{\rm K}\gtrsim300$~K can reproduce the high HCNH$^{+}$ abundances in the CMZ of NGC~253. 

When varying $n_{{\rm H}_{2}}$ in Fig.~\ref{Fig:chempl}b, the peak HCNH$^{+}$ abundances remain relatively constant, with differences within an order of magnitude. HCNH$^{+}$ reaches its peak abundance earlier at higher $n_{{\rm H}_{2}}$, as chemical timescales are inversely proportional to $n_{{\rm H}_{2}}$ for two-body reactions. 
In Fig.~\ref{Fig:chempl}c, we find that HCNH$^+$ abundances decrease with decreasing $A_{\rm v}$ at $A_{\rm v}\lesssim$5 due to photodissociation. At $A_{\rm v}\gtrsim$5, the HCNH$^{+}$ abundances are nearly independent of $A_{\rm v}$. In Fig.~\ref{Fig:chempl}d, the model results also show no dependence on the interstellar radiation field at $A_{\rm v}=20$ (see Table~\ref{Tab:phy}). The independence can be explained by the fact that regions with such high $A_{\rm v}$ are well shielded from ultraviolet radiation. 

Figure~\ref{Fig:chempl}e shows that higher cosmic-ray ionization rates result in lower HCNH$^{+}$ abundances, consistent with \citet{2021A&A...651A..94F}. This is because high cosmic-ray ionization rates can efficiently ionize H$_2$, leading to an enhancement of free electrons. These electrons can subsequently destroy molecular ions, including HCNH$^{+}$ and key precursors to its formation, via dissociative recombination reactions, thereby reducing the HCNH$^{+}$ abundance.

%Thus, elevated cosmic-ray ionization rates primarily reduce HCNH$^{+}$ abundances.

As shown above, the increased H$_{2}$ number densities, visual extinctions, interstellar radiations, and cosmic-ray ionization rates do not significantly increase the HCNH$^{+}$ abundances. Given the high gas temperatures in the CMZ of NGC~253 \citep{2019ApJ...871..170M,2020MNRAS.491.4573R,2023A&A...675A.151H,2024ApJ...961...18T}, we suggest that the unexpectedly high HCNH$^{+}$ abundances could result from chemical enhancement triggered by high gas temperatures. However, recent ALCHEMI studies show that the cosmic-ray ionization rates are very high, $\gtrsim 10^{3}\zeta_{0}$, in these GMCs \citep[e.g.,][]{2021ApJ...923...24H,2022ApJ...931...89H,2022ApJ...939..119B,2024ApJ...977...38B}. At such high cosmic-ray ionization rates, none of the static models can reproduce the high HCNH$^{+}$ abundances observed at their gas temperatures \citep[e.g., $T_{K}\lesssim$300~K;][]{2019ApJ...871..170M}. In contrast, chemically related species such as HCN and HNC can be successfully modeled with these static models \citep{2022ApJ...939..119B,2024ApJ...977...38B}. We also note that gas and dust may be not well coupled in low-density regions traced by HCNH$^{+}$. In these regions, dust temperatures are expected to be lower than gas temperatures, as confirmed in the CMZ of our Galaxy \citep[e.g.,][]{2013A&A...550A.135A,2016A&A...586A..50G}. Since gas evaporation from dust grain mantles is regulated by dust temperatures, lower dust temperatures would lead to even lower HCNH$^{+}$ abundances. Nevertheless, our main conclusion that the static models cannot reproduce the observed high HCNH$^{+}$ abundances remains unchanged.

\subsection{PDRs}\label{sec.mod2}
HCNH$^{+}$ transitions are likely to trace low-density regions (see Sect.~\ref{sec.density}), indicating that they may arise from PDRs. To explore this scenario, we used the \textit{Chempl} code for PDR model calculations \citep{2021RAA....21...77D}. As shown in Sect.~\ref{sec.mod1}, high gas temperatures can effectively increase the HCNH$^{+}$ abundances. In order to reproduce the observed HCNH$^{+}$ abundances, we used $T_{\rm k}=$500~K and a gas density of 10$^{3}$~cm$^{-3}$ as our fiducial parameters in the PDR models where dust temperatures are approximated following Eq.~(9.18) in \citet{2005pcim.book.....T}:
\begin{align} 
    T_{\rm d}^{5} & = 8.9\times 10^{-11}\nu_{0} G_{0}{\rm exp}[-1.8A_{\rm V}]+2.78^{5} \label{f:Td} \\
       &\quad +3.4\times 10^{-2} (0.42-{\rm ln}[3.5\times 10^{-2}\tau_{\rm 100\mu m}T_{0}])\tau_{\rm 100\mu m}T_{0}^{6}  \nonumber
\end{align}
where $\nu_{0}=3\times 10^{15}$, $\tau_{\rm 100\mu m}=10^{-3}$, and $T_{0}=12.2G_{0}^{1/5}$. The interstellar radiation field strength at the PDR edge, $G_0$, and the cosmic-ray ionization rate, $\zeta$, were the only parameters varied in our calculations.

\begin{figure}[!htbp]
\centering
\includegraphics[width = 0.49 \textwidth]{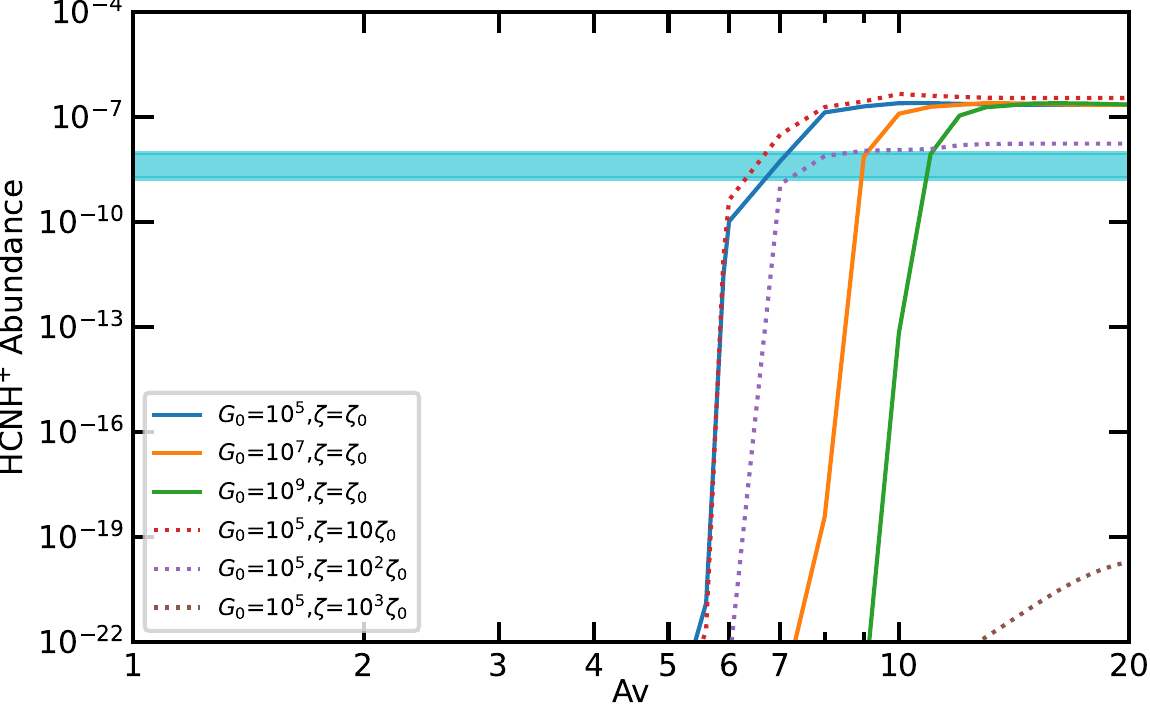}
\caption{{HCNH$^{+}$ abundance as a function of visual extinctions at different interstellar radiation fields and cosmic ray ionization rates, calculated from the PDR models of \textit{Chempl} \citep{2021RAA....21...77D}. The corresponding values are shown in the legend. The cyan-shaded region indicates the observed HCNH$^{+}$ abundances in this study. }\label{Fig:pdr}}
\end{figure}

The modeling results are shown in Figure~\ref{Fig:pdr}. As the visual extinction, $A_{\rm V}$, increases, HCNH$^+$ abundances generally increase for a given $G_{0}$ and $\zeta$. For different $G_{0}$, HCNH$^{+}$ abundances are nearly constant at high $A_{\rm V}$ (i.e., $A_{\rm V}\gtrsim 15$). This is because high interstellar radiation fields are likely to destroy HCNH$^{+}$ at low $A_{\rm V}$, making HCNH$^{+}$ most abundant in well-shielded regions. 
However, it should be noted that high values of $\zeta$ significantly reduce HCNH$^{+}$ abundances. While the observed HCNH$^{+}$ abundances can be reproduced for $\zeta \lesssim 10^{2}\zeta_{0}$, PDR models fail to match the observed values at the expected high cosmic-ray ionization rates, $\zeta \gtrsim 10^{3}\zeta_{0}$ \citep[e.g.,][]{2021ApJ...923...24H,2022ApJ...931...89H,2022ApJ...939..119B,2024ApJ...977...38B}.

\subsection{Shocks}\label{sec.mod3}
In Sect.~\ref{sec.mod1}, high kinetic temperatures of $T_{\rm K}\gtrsim300$~K are required to produce the observed high HCNH$^{+}$ abundances in environments associated with high cosmic-ray ionization rates (see Fig.~\ref{Fig:chempl}e). Such high kinetic temperatures can be caused by shock heating \citep{2023A&A...675A.151H}. Furthermore, \citet{2023A&A...675A.151H} find that shocked gas, traced by SiO transitions, has low H$_{2}$ number densities of $\lesssim 10^{4}$~cm$^{-3}$, similar to the density regime probed by HCNH$^{+}$ (see Sect.~\ref{sec.density}). This leads us to speculate that shocks in the CMZ may enhance HCNH$^{+}$ abundances. To investigate the impact of shocks on HCNH$^{+}$ abundances, we employed the UCLCHEM\footnote{\url{https://uclchem.github.io/}} code \citep{2017AJ....154...38H} to run shock models. 
In these models, we also augment the UMIST RATE12 chemical network with ${\rm C_{2}N^{+} + H_{2} } \to {\rm HCNH^{+} + C}$, as this reaction may contribute to the formation of HCNH$^{+}$ \citep{2023A&A...679A..39G}. This code involves two phases. In phase 1, we modeled a static cloud of isothermal gas with $T_{K}=$10~K and a gas density of 10$^{3}$~cm$^{-3}$ in atomic/ionic form with no molecules, leading to a model of a molecular cloud used in phase 2. In phase 2, we adopted C-type shocks for our calculations, parameterized according to \citet{2008A&A...482..549J}, similar to the approach in \citet{2023A&A...675A.151H}. In this approach, the preshock magnetic field strength was approximated using Eq.~(63) of \citet{1983ApJ...264..485D}.

\begin{figure}[!htbp]
\centering
\includegraphics[width = 0.49 \textwidth]{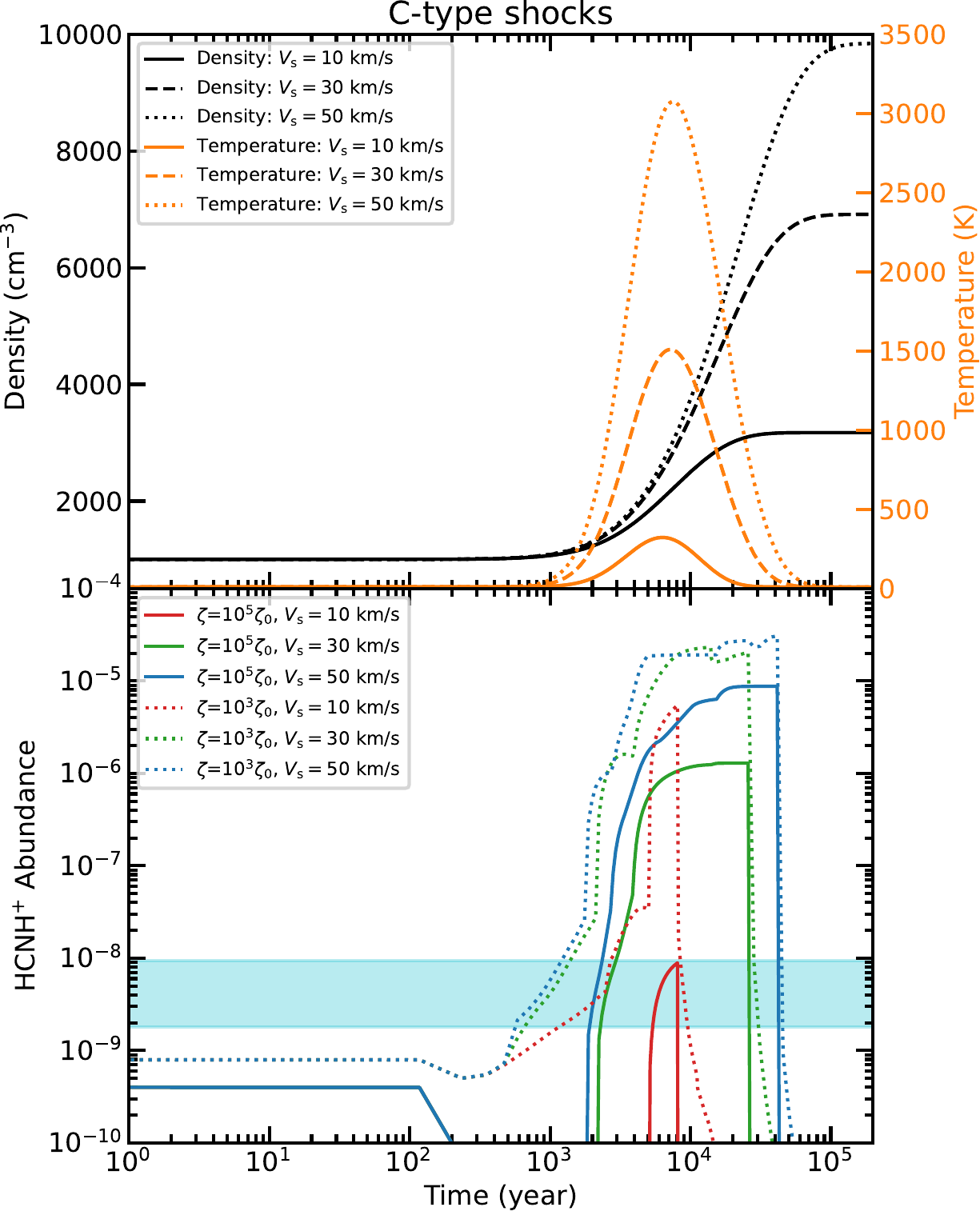}
\caption{{Physical structures and the evolution of HCNH$^{+}$ abundances in C-type shocks, calculated from the UCLCHEM models \citep{2017AJ....154...38H}. \textit{Top:} C-shock physical structures of our models for the three different shock velocities. The black and orange lines represent the density and temperature profiles of shock models. \textit{Bottom:} HCNH$^{+}$ abundances as a function of time for C-type shocks. The preshock gas density is fixed to 10$^{3}$~cm$^{-3}$ in the models. Different lines correspond to different cosmic ray ionization rates and shock velocities which are shown in the legend. The cyan-shaded region indicates the observed HCNH$^{+}$ abundances in this study. GIF animations of the shock models at a given $\zeta=10^{3}\zeta_{0}$ and $\zeta=10^{5}\zeta_{0}$ are available via \href{https://gongyan2444.github.io/img/c-shock-zeta1e3.gif}{link1} and \href{https://gongyan2444.github.io/img/c-shock-zeta1e5.gif}{link2}, respectively. }\label{Fig:shocks}}
\end{figure}

In the shock models, we mainly tested with different cosmic ray ionization rates and shock velocities with a fixed preshock gas density of 10$^{3}$~cm$^{-3}$. The modeling results are presented in Fig.~\ref{Fig:shocks}. In these shock models, the temperatures reach their maximum around the nearly identical timescale at $6\times 10^{3}$~yr. In this time evolution plot, an enhancement of HCNH$^{+}$ abundances is evident around the timescales close to the time of the maximum temperature and the HCNH$^{+}$ abundances are significantly lower than the observed values before and after the shock phase. Both cosmic ray ionization rates and shock velocities affect HCNH$^{+}$ abundances. In these models, elevated cosmic ray ionization rates are necessary to reproduce the observed HCNH$^{+}$ abundance enhancements. With $\zeta=\zeta_{0}$, shock models cannot account for the observed HCNH$^{+}$ abundance. Only by increasing $\zeta$ can the observed HCNH$^{+}$ levels be achieved. This is because cosmic ray ionization rates enhance the abundances of molecular ions, which in turn led to the HCNH$^{+}$ enhancement. However, when $\zeta \gtrsim 10^{3} \zeta_{0}$, the modeled HCNH$^{+}$ abundances decrease.  This occurs because high cosmic ray ionization rates efficiently ionize H$_{2}$ and produce the elevated electron abundances, which in turn efficiently destroy molecular ions. In contrast, higher shock velocities lead to significantly elevated HCNH$^{+}$ abundances and sustain these enhanced levels over longer timescales. 

We also compared the HCN and HNC abundances derived from \citet{2024ApJ...977...38B} with our shock models. To reproduce the observed HCN and HNC abundances, higher cosmic ray ionization rates of of $\gtrsim 10^{3} \zeta_{0}$ are required, consistent with the findings of previous studies \citep{2022ApJ...939..119B,2024ApJ...977...38B}. However, it is important to note that the densities probed by HCN and HNC \citep{2024ApJ...977...38B} appear to be higher than those traced by HCNH$^{+}$, suggesting that HCNH$^{+}$ may trace a distinct component of the molecular gas in the CMZ of NGC253 compared to HCN and HNC.

While our static (Section~\ref{sec.mod1}) and PDR (Section~\ref{sec.mod2}) models failed to reproduce the observed HCNH$^{+}$ abundances at high cosmic ray ionization rates of $\zeta \sim 10^{5}\zeta_{0}$, our C-type shock models can reproduce the high HCNH$^{+}$ abundances at such extreme physical conditions with shock velocities of $\varv_{\rm s}\gtrsim$10~\kms. However, we note that a single shock event cannot sustain high HCNH$^{+}$ abundances due to the rapid destruction of HCNH$^{+}$ molecules in post-shock gas. Previous studies suggest that the CMZ may undergo multiple shock episodes, potentially triggered by outflows, star formation, and cloud-cloud collisions \citep{2023A&A...675A.151H}. These recurrent shocks could, in principle, sustain the elevated HCNH$^{+}$ abundances observed. We therefore propose that HCNH$^{+}$ primarily originates from shocked gas in the CMZ.

\subsection{Formation and destruction paths}
In the adopted chemical network, HCNH$^{+}$ can be produced through ion-neutral reactions as follows:
\begin{equation}\label{f.form1}
\begin{split}
{\rm H_{3}^{+} + HCN/HNC} &\to {\rm HCNH^{+} + H_{2}} \;
\end{split}
\end{equation}
\begin{equation}\label{f.form2}
\begin{split}
{\rm HCO^{+} + HCN/HNC} &\to {\rm HCNH^{+} + CO} \;
\end{split}
\end{equation}
\begin{equation}\label{f.form3}
\begin{split}
{\rm C^{+} + NH_{3}} &\to {\rm HCNH^{+} + H} \;
\end{split}
\end{equation}
\begin{equation}\label{f.form4}
\begin{split}
{\rm H_{2} + HCN^{+}/HNC^{+}} &\to {\rm HCNH^{+} + H} \;
\end{split}
\end{equation}
\begin{equation}\label{f.form5}
\begin{split}
{\rm C_{2}N^{+} + H_{2} } &\to {\rm HCNH^{+} + C} \;.
\end{split}
\end{equation}
while its destruction occurs via the following reactions, 
\begin{equation}\label{f.des1}
\begin{split}
{\rm HCNH^{+} + e^{-}} &\to {\rm HCN + H} \\
 &\to {\rm HNC + H} \\
 &\to {\rm CN + H + H}  \;.
\end{split}
\end{equation}
\begin{equation}\label{f.des2}
\begin{split}
{\rm CH + HCNH^{+}}  &\to {\rm HCN/HNC + CH_{2}^{+}} \\
\end{split}
\end{equation}
\begin{equation}\label{f.des3}
\begin{split}
{\rm CH_{2} + HCNH^{+}}  &\to {\rm HCN/HNC + CH_{3}^{+}} \\
\end{split}
\end{equation}
\begin{equation}\label{f.des4}
\begin{split}
{\rm NH_{3} + HCNH^{+}}  &\to {\rm HCN/HNC + NH_{4}^{+}} \\
\end{split}
\end{equation}

In cold regions ($T_{\rm K}\sim$10~K), reactions~(\ref{f.form1})--(\ref{f.form3}) were thought to dominate the formation of HCNH$^{+}$, while dissociative recombination through reaction~(\ref{f.des1}) serves as the main destruction pathway \citep[e.g.,][]{1991A&A...247..487S}. However, both HCN and HNC act as reactants and products, presenting a ``chicken-and-egg problem". Chemical modeling by \citet{2021A&A...651A..94F} suggests that reactions~(\ref{f.form3}) and (\ref{f.form4}) are the primary formation pathways, with HCN$^{+}$ mainly produced via CN + H$_{3}^{+}$, effectively resolving the chicken-and-egg issue. Reaction~(\ref{f.form5}) was previously overlooked because it was not included in the UMIST RATE12 chemical network. However, using the rate coefficient of reaction~(\ref{f.form5}) \citep{Knight1988,2014MNRAS.437..930L}, \citet{2023A&A...679A..39G} compared the chemical reaction rates of reactions~(\ref{f.form1})--(\ref{f.form5}) and suggested that reaction~(\ref{f.form5}) could make non-negligible contributions. \citet{2021A&A...651A..94F} also explored the chemistry of warm regions ($T_{\rm K} \sim 27$ K), finding that reaction~(\ref{f.form2}) becomes the primary formation mechanism, while the dominant destruction pathway remains unchanged.

\begin{figure}[!htbp]
\centering
\includegraphics[width = 0.49 \textwidth]{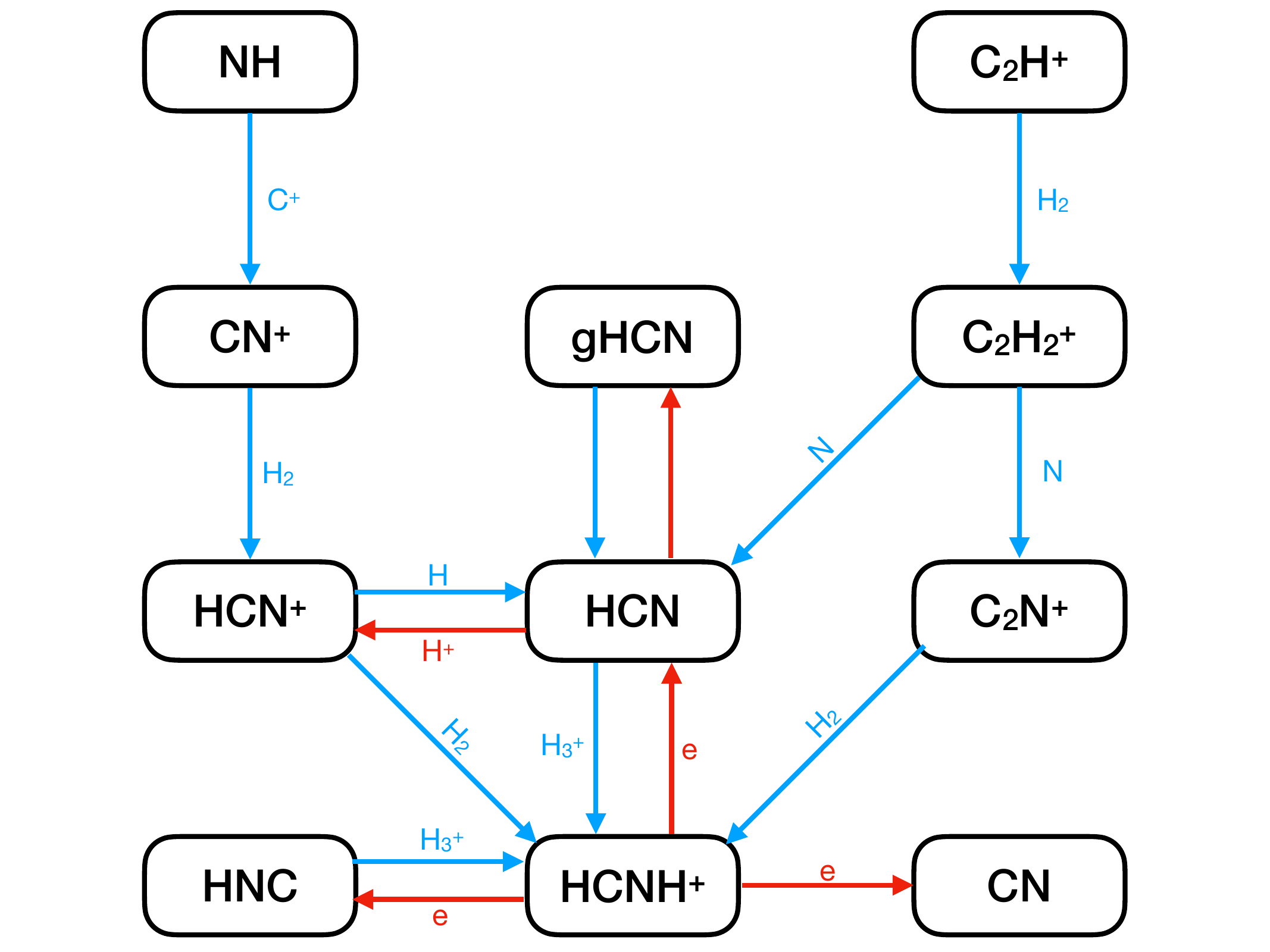}
\caption{{Schematic diagram of the simplified chemical network that dominates the HCNH$^{+}$ chemistry in the shock model.}\label{Fig:diagram}}
\end{figure}

We also explored the chemistry based on the shock models (see Sect.~\ref{sec.mod3}). The HCNH$^{+}$ chemistry in the shock models is presented in Fig.~\ref{Fig:diagram}. For instance, we employed a shock model with a cosmic-ray ionization rate $\zeta=10^{5}\zeta_{0}$ and a shock velocity of $\varv_{\rm s}$=30~\kms. In this model, the primary formation pathway for HCNH$^+$ during the early phase of C-type shocks (at timescales of $\lesssim 1.4 \times 10^{4}$ years in Fig.~\ref{Fig:shocks}) is reaction~(\ref{f.form4}) involving H$_{2}$+HCN$^{+}$, similar to the cold model in \citet{2021A&A...651A..94F}. As the shock evolves into the second phase (at timescales of $(1.4-2) \times 10^{4}$ years in  Fig.~\ref{Fig:shocks}), reaction~(\ref{f.form5}) becomes the dominant formation pathway, contributing to $\gtrsim$80\% of the HCNH$^+$ formation rate. Eventually, at later stages, reaction~(\ref{f.form4}) once again takes over as the primary formation mechanism. Additionally, reaction~(\ref{f.form1}) via H$_{3}^{+}$ and HCN also significantly contributes to the formation of HCNH$^{+}$ molecules during the early phase of shocks (at timescales of $5 \times 10^3$ to $1.4 \times 10^4$ years; Fig.~\ref{Fig:shocks}), accounting for 16\%--27\% of the HCNH$^+$ formation rate during this period.

Another factor contributing to the elevated HCNH$^+$ abundances is the temperature dependence of the destruction reaction~(\ref{f.des1}). The rate coefficient for this reaction decreases with increasing gas temperature, following a power-law dependence of $-0.65$ \citep{2013A&A...550A..36M}. At a gas temperature of 300~K, the reaction rate coefficient is about an order of magnitude lower than at 10~K. For demonstration, we once again employed the shock model with a cosmic-ray ionization rate $\zeta=10^{5}\zeta_{0}$ and a shock velocity of $\varv_{\rm s}$=30~\kms. During the shock phase (on timescales of $2 \times 10^{3}$ to $3 \times 10^{4}$ years in Fig.~\ref{Fig:shocks}), reactions~(\ref{f.des2})--(\ref{f.des4}) become increasingly efficient, but the overall destruction of HCNH$^+$ is significantly reduced. Over the entire process, reaction~(\ref{f.des1}) remains the dominant mechanism for the destruction of HCNH$^+$. Thus, high temperatures induced by shock heating substantially mitigate the destruction of HCNH$^+$, leading to an enhancement in its abundance. 

%Furthermore, reactions~(\ref{f.des1})--(\ref{f.des4}) exhibit nearly identical branching ratios for producing HCN and HNC. When HCNH$^{+}$ abundances are high, these reactions result in comparable amounts of HCN and HNC, contributing to the low HCN/HNC abundance ratios observed in the CMZ of NGC~253 \citep{2022ApJ...939..119B,2024ApJ...977...38B}. Therefore, HCNH$^{+}$ likely plays a significant role in shaping the low HCN/HNC ratios in environments like the CMZ of NGC~253.

\section{Summary and conclusions}\label{Sec:sum}
To explore the behavior and significance of HCNH$^{+}$ under extreme conditions, we analyzed four rotational transitions of HCNH$^{+}$ from the ALCHEMI large program, focusing on the physical and chemical properties of giant molecular clouds in the heart of the starburst galaxy NGC~253. Our main results are summarized as follows:
\begin{itemize}

    \item[1.] HCNH$^{+}$ emission is widespread with a radial extent of $\sim$300~pc in the central molecular zone, effectively tracing large-scale gas structures within the CMZ of NGC~253.\\ 

    \item[2.] Our non-LTE analysis yields HCNH$^{+}$ column densities of (0.3--1.2)$\times 10^{15}$~cm$^{-2}$, with corresponding HCNH$^{+}$ abundances relative to H$_{2}$ of (1.8--9.4)$\times 10^{-9}$. Most of these HCNH$^{+}$ abundances are higher than all reported values in the Milky Way. This highlights the elevated HCNH$^{+}$ abundances in the CMZ of NGC~253. Furthermore, we observed a trend of decreasing HCNH$^{+}$ abundances in GMCs as they approach the center of the CMZ. \\

    \item[3.] Using the quantum mechanical coupled states approximation, we calculated collisional rate coefficients for HCNH$^{+}$ induced by {\it para}-H$_2$ and {\it ortho}-H$_2$ at kinetic temperatures up to 500 K. Incorporating these new coefficients into a non-LTE modeling framework and applying Monte Carlo Markov chain analysis, we suggest that HCNH$^{+}$ emissions originate from regions with H$_2$ number densities of $\sim 10^{2.80}-10^{3.55}$~cm$^{-3}$ with typical errors of $\sim$0.2~dex, confirming HCHN$^{+}$ as a tracer of low-density regions, even in nuclear starburst environments. \\

    \item[4.] Astrochemical calculations assuming static, PDR, and shock environments have been performed. These modeling results indicate that the unexpectedly high HCNH$^{+}$ abundances observed in the CMZ of NGC~253 can be attributed to chemical enhancement, driven by elevated gas temperatures and cosmic ray ionization rates in shocked, low-density gas. \\
    
\end{itemize}

The detection of HCNH$^{+}$ in the CMZ of NGC~253 reveals significant differences compared to its presence in quiescent molecular clouds in the Solar neighborhood. This study opens new avenues for understanding the complex chemistry and physical conditions in extreme environments of starburst galaxies. Given its role as a tracer of low-density regions, future investigations could benefit from mapping HCNH$^{+}$ across various galactic environments to explore its potential as a diagnostic tool for studying various stages of molecular cloud evolution, the early stages of star formation, and the impacts of star formation feedback.

\begin{acknowledgement}\label{sec.ack}
We acknowledge the ALMA staff for their assistance with our observations. Y.G. is supported by the Strategic Priority Research Program of the Chinese Academy of Sciences, Grant No. XDB0800301. C. Henkel acknowledges support by the Chinese Academy of Sciences President's International Fellowship Initiative under grant No.~2025PVA0048. We also acknowledge financial support from the European Research Council (Consolidator Grant COLLEXISM, Grant Agreement No. 811363) and the support from CEA/GENCI for granting us access to the TGCC/IRENE supercomputer under the A0110413001 project. F.J.D. is supported by National Key R\&D Program of China grant 2023YFA1608000 and NSFC grant 12041305. S.V and M.B acknowledge the support from the European Research Council (ERC) Advanced Grant MOPPEX 833460. X.D.Tang acknowledges the support of the National Key R\&D Program of China under grant No.~2023YFA1608002, the Chinese Academy of Sciences (CAS) ``Light of West China" Program under grant No.~xbzg-zdsys-202212, the Tianshan Talent Program of Xinjiang Uygur Autonomous Region under grant No.~2022TSYCLJ0005, and the Natural Science Foundation of Xinjiang Uygur Autonomous Region under grant No.~2022D01E06. W.Y. acknowledges the support from the National Natural Science Foundation of China (12403027), China Postdoctoral Science Foundation (2024M751376), and Jiangsu Funding Programme for Excellent Postdoctoral Talent (2024ZB347). This research has made use of spectroscopic and collisional data from the EMAA database (https://emaa.osug.fr and https://dx.doi.org/10.17178/EMAA). EMAA is supported by the Observatoire des Sciences de l’Univers de Grenoble (OSUG). This research has made use of NASA's Astrophysics Data System. This work also made use of the Cube Analysis and Rendering Tool for Astronomy (CARTA) software \citep{2021zndo...3377984C} and Python libraries including Astropy\footnote{\url{https://www.astropy.org/}} \citep{2013A&A...558A..33A}, NumPy\footnote{\url{https://www.numpy.org/}} \citep{5725236}, SciPy\footnote{\url{https://www.scipy.org/}} \citep{jones2001scipy}, Matplotlib\footnote{\url{https://matplotlib.org/}} \citep{Hunter:2007}. This article makes use of the following ALMA data: ADS/JAO.ALMA\#2017.1.00161.L and ADS/JAO.ALMA\#2018.1.00162.S. ALMA is a partnership of ESO (representing its member states), NSF (USA) and NINS (Japan), together with NRC (Canada), NSTC and ASIAA (Taiwan), and KASI (Republic of Korea), in cooperation with the Republic of Chile. The Joint ALMA Observatory is operated by ESO, AUI/NRAO and NAOJ. The National Radio Astronomy Observatory is a facility of
the National Science Foundation operated under cooperative agreement by Associated Universities, Inc. During the review process, we were deeply saddened by the passing of our esteemed colleague, Prof. Karl M. Menten, whose contributions to this work and to the field will always be remembered. We thank the anonymous referee for helpful comments that improve this manuscript. 
\end{acknowledgement} 

\bibliographystyle{aa}
\bibliography{references}

\begin{appendix}
\onecolumn
\renewcommand\thefigure{\thesection.\arabic{figure}}    
\setcounter{figure}{0}
\section{Assessing errors from coupled state rate coefficients in radiative transfer calculations}
\label{appendix_Tr_ratio}
\begin{figure}[hb]
    \centering
    \includegraphics[width = 0.48\linewidth, trim = 0 0 0 10, clip = true]{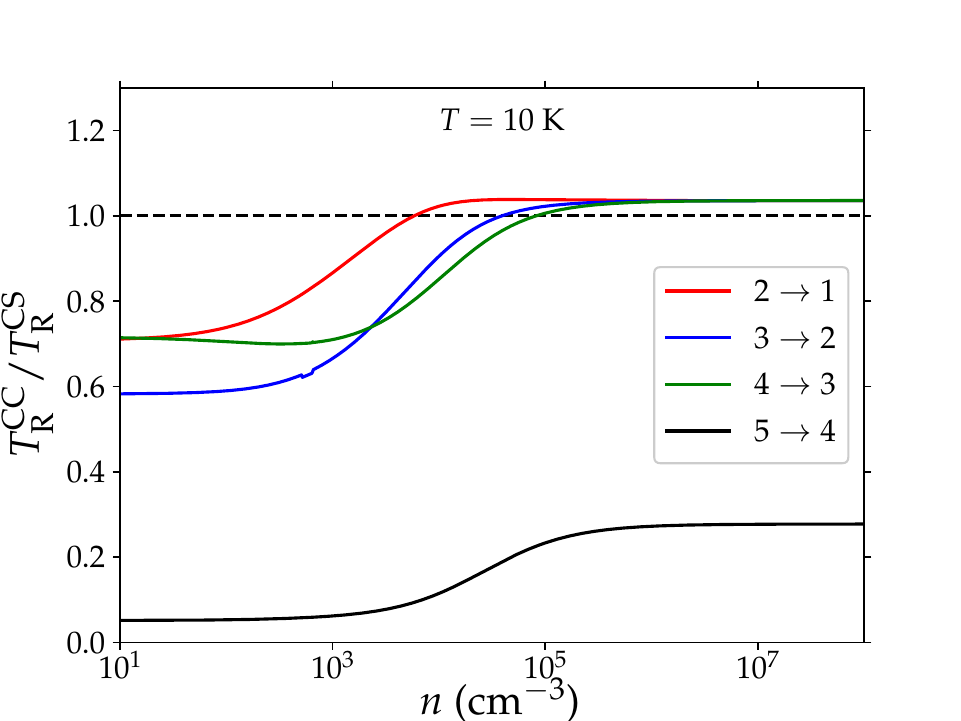}
    \includegraphics[width = 0.48\linewidth, trim = 0 0 0 10, clip = true]{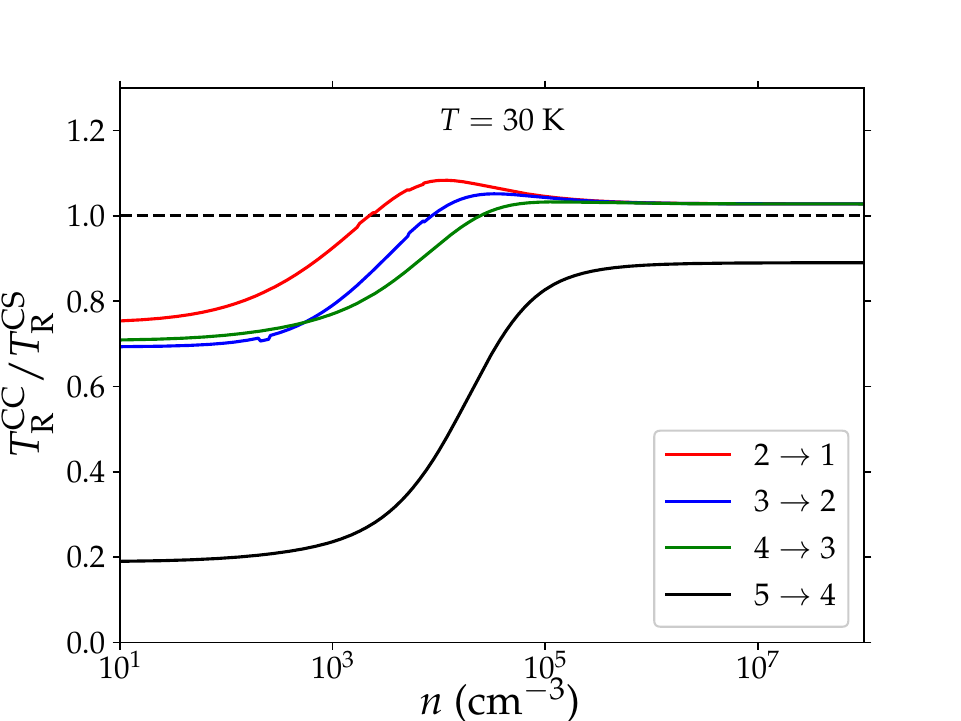}
    \caption{Dependence of radiation temperature ratios on gas density for the $1\to0$, $2\to1$, $3\to2$, and $4\to3$ emission lines. Superscripts CC and CS denote that the rate coefficients used in the radiative transfer calculations were obtained using the close-coupling (CC) method and the coupled states (CS) approximation, respectively. The dashed line indicates perfect agreement between the methods. All calculations are performed under the LVG formalism as implemented in the \texttt{RADEX} code, assuming a FWHM line width of 1.0~\kms\,and an HCNH$^{+}$ column density of $2.3\times10^{13}$~cm$^{-2}$ \citep{2007A&A...468..627V}.}
    \label{fig:Tr_ratio}
\end{figure}
\newpage
\section{Systematic comparison of HCNH$^+$ rate coefficients induced by collisions with {\it o}- and {\it p}-H$_2$}
\label{appendix_rates}
\begin{figure}[hb]
    \centering
    \includegraphics[width = 0.48\linewidth, trim = 0 10 0 10, clip = true]{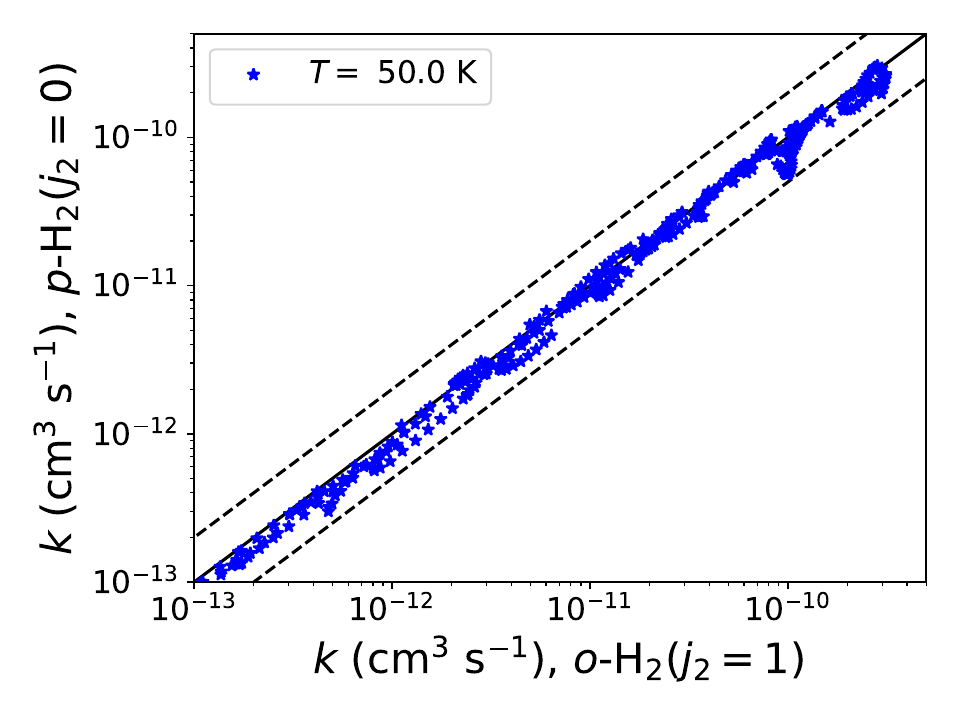}
    \includegraphics[width = 0.48\linewidth, trim = 0 10 0 10, clip = true]{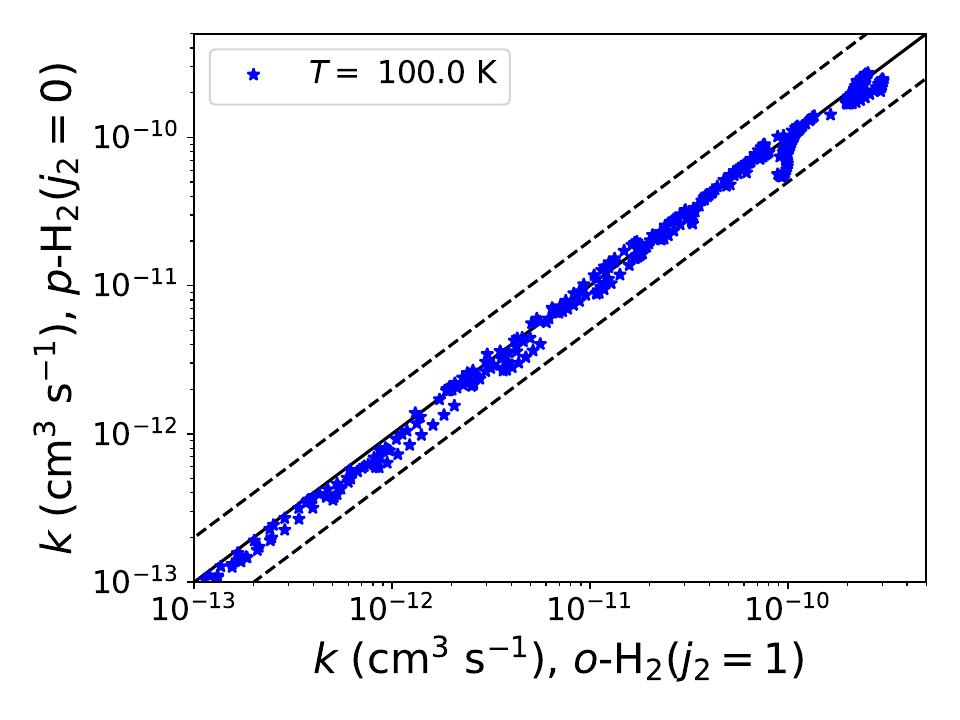}
    \includegraphics[width = 0.48\linewidth, trim = 0 10 0 10, clip = true]{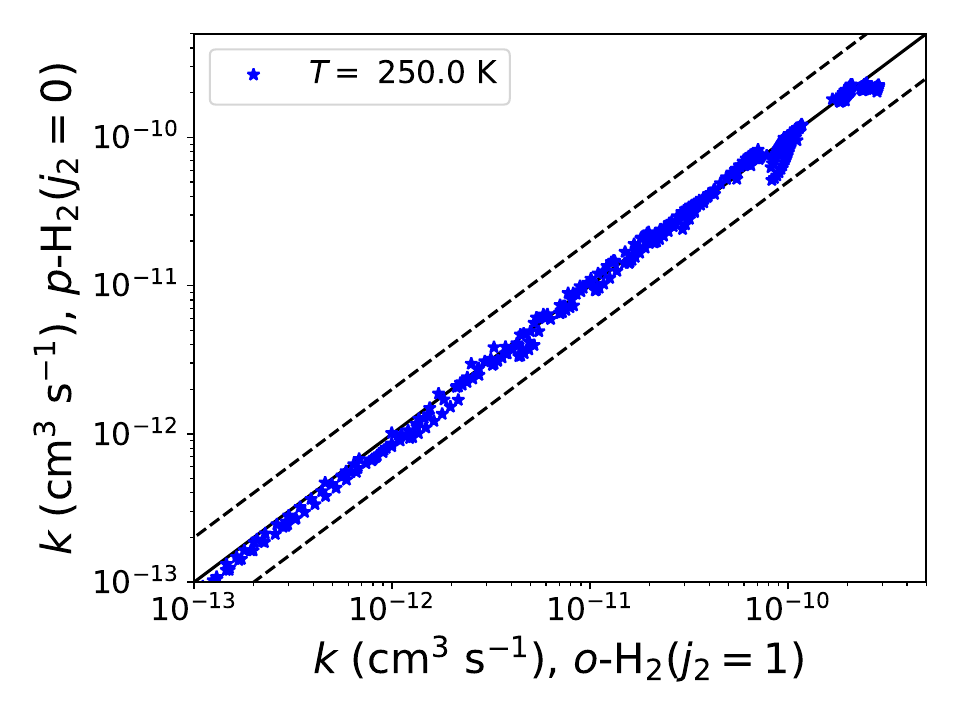}
    \includegraphics[width = 0.48\linewidth, trim = 0 10 0 10, clip = true]{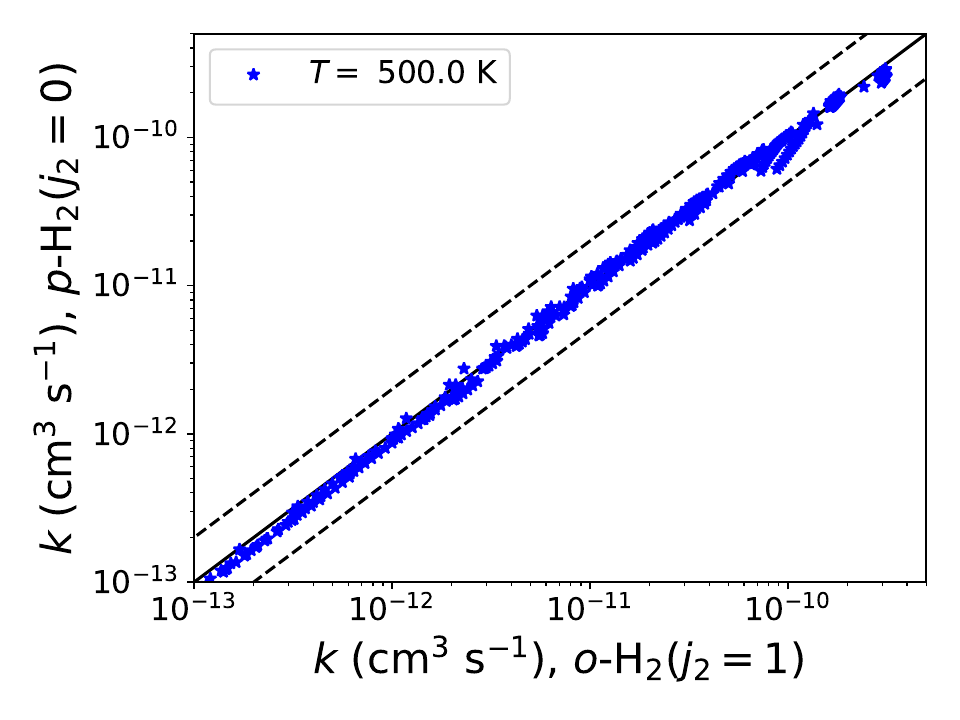}
    \caption{Comparison of HCNH$^+$ rate coefficients induced by collisions with {\it o}- and {\it p}-H$_2$ for selected temperatures. The x- and y-axes represent the {\it o}- and {\it p}-H$_2$ scattering data, respectively. The dashed lines indicate an agreement region within a factor of 2.0.}
    \label{fig:ortho_para-T_50-500}
\end{figure}
\end{appendix}

\end{document}